\documentclass[aps,prb,twocolumn,superscriptaddress,floatfix,citeautoscript,hyperref=pdftex]{revtex4-1}
\usepackage{times}
\usepackage{graphicx}
\usepackage{dcolumn}
\usepackage{bm}
\usepackage{color}
\usepackage{amsmath}
\usepackage{amssymb}
\newcommand {\beq} {\begin{equation}}
\newcommand {\eeq} {\end{equation}}
\newcommand {\bqa} {\begin{eqnarray}}
\newcommand {\eqa} {\end{eqnarray}}

\usepackage[colorlinks=true]{hyperref}

\begin{document}

\title{Disorder-robust phase crystal in high-temperature superconductors stabilized by strong correlations}

\author{Debmalya Chakraborty}
\affiliation{Department of Physics and Astronomy, Uppsala University, Box 516, S-751 20 Uppsala, Sweden}

\author{Tomas L\"{o}fwander}
\affiliation{Department of Microtechnology and Nanoscience - MC2,Chalmers University of Technology, SE-412 96 G\"{o}teborg, Sweden}

\author{Mikael Fogelstr\"{o}m}
\affiliation{Department of Microtechnology and Nanoscience - MC2,Chalmers University of Technology, SE-412 96 G\"{o}teborg, Sweden}

\author{Annica M. Black-Schaffer}
\affiliation{Department of Physics and Astronomy, Uppsala University, Box 516, S-751 20 Uppsala, Sweden}

\begin{abstract}

The simultaneous interplay of strong electron-electron correlations, topological zero-energy states, and disorder is yet an unexplored territory but of immense interest due to their inevitable presence in many materials. Copper oxide high-temperature superconductors (cuprates) with pair breaking edges host a flat band of topological zero-energy states, making them an ideal playground where strong correlations, topology, and disorder are strongly intertwined. Here we show that this interplay in cuprates generates a new phase of matter: a fully gapped ``phase crystal" state that breaks both translational and time-reversal invariance, characterized by a modulation of the $d$-wave superconducting phase co-existing with a modulating extended $s$-wave superconducting order. In contrast to conventional wisdom, we find that this phase crystal state is remarkably robust to omnipresent disorder, but only in the presence of strong correlations, thus giving a clear route to its experimental realization.

\end{abstract}

\maketitle

Different phases of matter appear in condensed matter physics due to the presence of strong correlations \cite{Lee06,Paschen21}, disorder \cite{Anderson58,Goldman98}, or topology \cite{Moore10,Sato17}. While they have all been intensively studied individually and also to some extent pair-wise \cite{Punnoose05,Meier18,Shi21}, the combination of all three effects is one of the currently most challenging problem in physics. Cuprates with pair breaking edges provide an interesting platform where this three-way interplay can become manifest. First, the $d$-wave pairing symmetry in cuprates leads to Andreev bound states at the pair breaking [110] edges forming flat bands of zero-energy states \cite{Kashiwaya00,Lofwander01}, protected by the bulk topology \cite{Ryu02,Sato11}. Secondly,  electron correlations are exceptionally strong in the cuprates, with the undoped parent compound even being a Mott insulator \cite{Lee06}. Finally, disorder, both intrinsic or extrinsic, is prominent in all cuprates \cite{Howald01,Alloul09}. It has already been shown that the flat band of zero-energy states is thermodynamically unstable due to the extensive ground-state degeneracy and is hence highly susceptible to electronic correlations \cite{Potter14}. However, the question of how these zero-energy states respond to a simultaneous presence of strong correlations and disorder is still an unsolved problem.

The thermodynamic instability of the zero-energy states opens for a possible appearance of competing phases, breaking the topological protection \cite{Honerkamp00,Lofwander00,Potter14,Black-Schaffer13,Nagai17,Matsubara20}. The likely most promising intrinsic scenario involves an exotic state called ``phase crystal", which spontaneously breaks both time-reversal and translational symmetries along the [110] edge \cite{Hakansson2015, Holmvall18, Holmvall20, Wennerdal20}. In this state, the phase of the $d$-wave superconducting order parameter is modulated along the edge, creating superflow patterns which lower the free energy by Doppler shifting some, but not all, states away from zero energy. This state was originally found in quasiclassical calculations \cite{Hakansson2015}, but subsequently also in a weak-coupling tight-binding model \cite{Wennerdal20}. 
However, these and most other studies of competing phases, do not take into account two of the most essential ingredients of cuprates: strong correlations and disorder. It is both unknown if the phase crystal even survives in the presence of strong correlations, and, most importantly, disorder is known to essentially eliminate the ground state degeneracy \cite{Kalenkov04,Ikegaya17}, thus making any competing phase seemingly unlikely when real-world disorder is present.

Here we perform fully self-consistent calculations of the superconducting state in cuprate superconductors and find that strong correlations stabilize a phase crystal state along the [110] edge that develops a full energy gap due to a co-existing $s$-wave superconducting order. We first show how strong correlations increase the number of zero-energy states for any uniform superconducting phase, contradicting simple topological arguments. Then, when allowing for a non-uniform solution, we find a cascade of phase transitions occurring at different temperatures: $d$-wave superconductivity occurs below a transition temperature $T_c$, the phase crystal appears further below at temperature $T^*\sim0.2 T_c$, and finally an additional extended $s$-wave order, with the same spatial modulations as the phase crystal, is generated below $T_s$, generating a full energy gap. Taken together, these phase transitions explain a set of so far seemingly contradictory experimental results \cite{Covington97,Dagan01,Gustafsson13,Alff97,Neils02}. Furthermore, we find that the phase crystal state is unexpectedly very robust to disorder, but notably only in the presence of strong correlations. Our results show that the combined effects of strong correlations and topology lead to the emergence of novel phases of matter that survives strong disorder in a highly non-intuitive manner.

\begin{figure*}[t]
\includegraphics[width=0.8\linewidth]{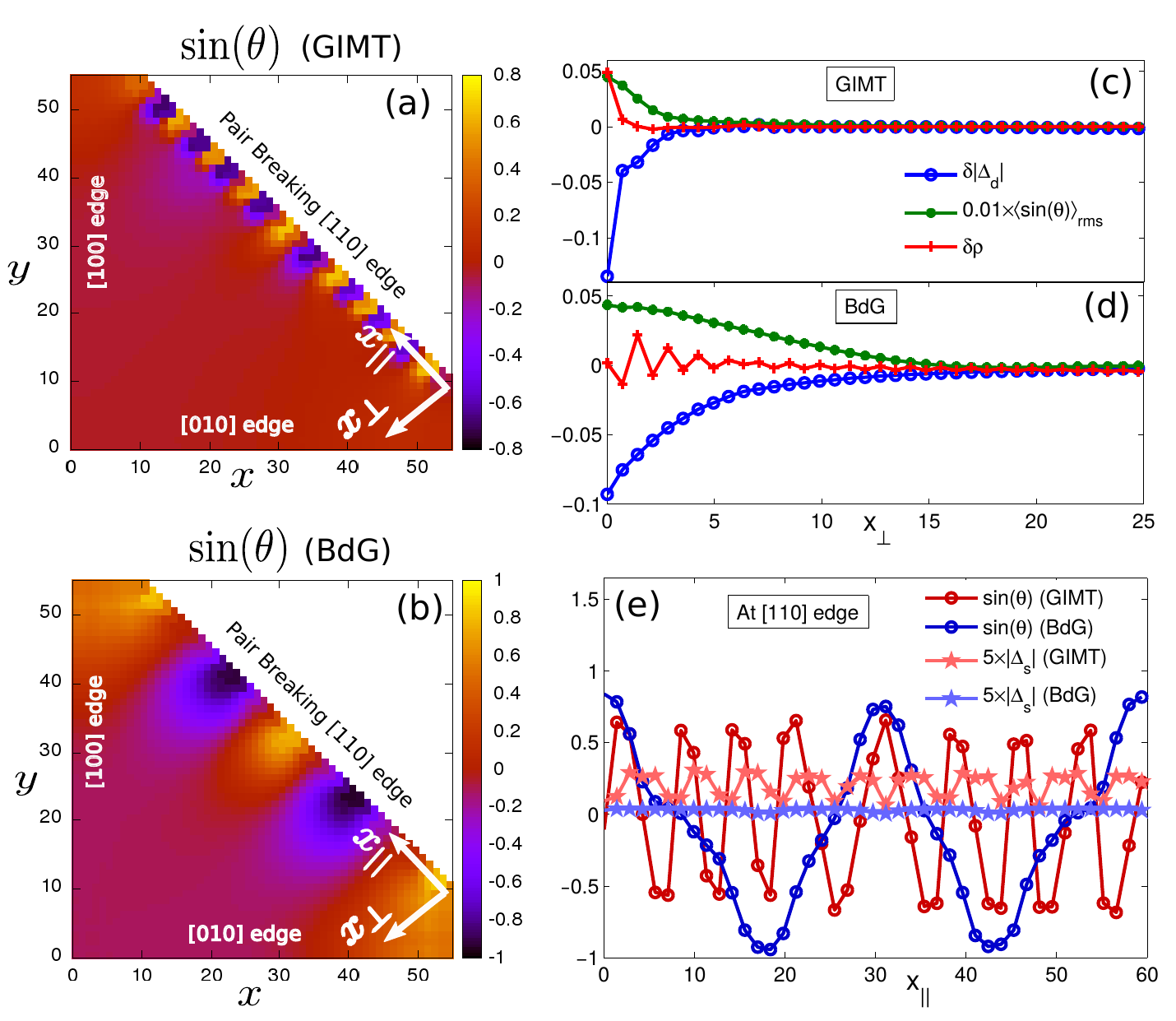} \caption{{\bf {Phase crystals for clean sample.}} Color density maps of the phase $\sin(\theta)$ of the local $d$-wave superconducting pairing amplitude within GIMT ({\bf{a}}) and BdG ({\bf{b}}) at temperature $T=0.01$. $x_{\parallel}$ and $x_{\perp}$ denote  distances along and perpendicular to the pair breaking [110] edge, respectively. ({\bf{c}},{\bf{d}}) Relative change in the magnitude of the $d$-wave pairing amplitude, defined as $\delta \left| \Delta_{d} \right|=\langle \left|\Delta_{d}\right|\rangle_{\textrm{rms}}-\left|\Delta_{d}(\textrm{bulk})\right|$, phase $\langle \textrm{sin}(\theta) \rangle_{\textrm{rms}}$, and relative change in the charge density defined as $\delta \rho =\langle \rho\rangle_{\textrm{rms}}-\rho(\textrm{bulk})$ plotted as a function of $x_{\perp}$ for GIMT and BdG, respectively. Here $\langle ... \rangle_{\textrm{rms}}$ denotes the root mean square average over all the sites in the $x_{\parallel}$ direction for a fixed $x_{\perp}$. 
({\bf{e}}) Line plots of the phase modulations together with the magnitude of the extended $s$-wave pairing amplitude $\left|\Delta_{s}\right|$ at the pair breaking edge.
See Methods for remaining parameters.}
\label{fig:cleanPC} 
\end{figure*}

{\noindent {\textbf {Model and approach.}}} The $d$-wave superconducting state in the strongly correlated cuprates has long been assumed to be described by the strong coupling repulsive Hubbard-$U$ model \cite{Scalapino95} and the equivalent $t$-$J$ model \cite{Anderson04}:
\begin{eqnarray}
{\cal H}_{\rm t-J}&=&\sum_{ij \sigma}t_{ij} {\cal P} \left(c^{\dagger}_{i \sigma} c_{j \sigma}+\textrm{H.c.}\right) {\cal P} - \sum_{i} \mu n_{i} \nonumber \\
&+&\sum_{\langle ij\rangle}J \left(\mathbf{S}_i \cdot \mathbf{S}_j-\frac{n_i n_j}{4}\right).
\label{eq:tJ}
\end{eqnarray}
Here $c_{i \sigma}^{\dagger}$ ($c_{i \sigma}$) is the creation (annihilation) operator of an electron with spin $\sigma$ at lattice site $i$ in a two-dimensional square lattice, $\mathbf{S}_i$ and $n_{i}$ are the spin and electron density operators, respectively, $\mu$ is the chemical potential fixing the average electron density, $\langle ij \rangle$ denotes nearest neighbor bonds, $J = 4t^2/U$ is the super-exchange interaction with $U$ being the onsite Hubbard repulsion strength, and $t_{ij}$ is the hopping amplitude for an electron from site $i$ to $j$, where we consider $t_{ij}=-t$ for nearest neighbor bonds and $t_{ij}=t^{\prime}$ for next-nearest neighbor bonds. For simplicity we express energies in units of $t$, lengths in units of the lattice spacing $a$, and set $\hbar=k_B=1$. Furthermore, ${\cal P}$ is the projection operator which prohibits double occupancies on each lattice site due to the strong onsite repulsive $U$. The effects of the projection operator can be implemented by the Gutzwiller approximation, where the Hamiltonian parameters are renormalized by statistical factors ensuring no double occupancy \cite{Zhang88}. The Gutzwiller approximation has been verified to agree well with variational Monte Carlo calculations, which treat the effects of the projection exactly \cite{Paramekanti01}, in both homogeneous systems \cite{Sensarma07,Fukushima08} and for bulk impurities \cite{Fukushima09}. In the presence of inhomogeneities, such as edges or disorder, the Gutzwiller approximation should be implemented with renormalization factors determined by the local electron density \cite{Garg08}. In this way, the Gutzwiller approximation becomes a powerful theoretical tool encapsulating the simultaneous effects of strong correlations and inhomogeneities. 

We perform a Gutzwiller inhomogeneous mean-field theory (GIMT) treatment of the Hamiltonian in Eq.~\eqref{eq:tJ} \cite{Garg08,Chakraborty14}. The super-exchange $J$ term gives rise to superconductivity with spin-singlet Cooper pairs living on nearest neighbor bonds, which can be represented by the local $d$- and extended $s$-wave pairing amplitudes \cite{Garg08}: $\Delta_{d}(i)=0.25(\Delta_{i}^{+x}+\Delta_{i}^{-x}-\Delta_{i}^{+y}-\Delta_{i}^{-y})$ and $\Delta_{s}(i)=0.25(\Delta_{i}^{+x}+\Delta_{i}^{-x}+\Delta_{i}^{+y}+\Delta_{i}^{-y})$. Here $\Delta_i^\delta$ is the pairing amplitude on the nearest neighbor bond in direction $\delta$.
We are further interested in the phase $\theta$ of the $d$-wave pairing amplitude characterized through $\Delta_d(i)=\left|\Delta_d(i)\right|\exp(i\theta)$. To quantify the effects of strong correlations, we compare our GIMT results with the results of standard weak-coupling Bogoliubov-de Gennes (BdG) calculations \cite{Zhubook}, where the important projection ${\cal P}$ in Eq.~\eqref{eq:tJ} is ignored and the only effect of the correlations is to create superconducting pairing. Finally, we introduce generic non-magnetic disorder by studying
\begin{equation}
{\cal H}={\cal H}_{\rm t-J}+\sum_{i} V_i n_{i},
\label{eq:tJdis}
\end{equation}
where $V_i$ is a site-dependent non-magnetic impurity potential drawn from a random distribution, such that $V_{i} \in [-V/2,V/2]$ uniformly, also known as Anderson disorder. 
Details of the GIMT and BdG methods, system geometry, and choice of the parameters are given in the Methods section.

{\noindent {\textbf {Clean superconductor edge.}}} We begin by looking at a clean superconductor with a pair breaking [110] edge. In Fig.~\ref{fig:cleanPC}(a,b) we show the phase $\theta$ of the $d$-wave pairing amplitude of the ground state at a low temperature, obtained with strong correlations within GIMT and as a comparison without strong correlations within BdG. In both cases $\theta$ acquires distinct and modulating non-zero values near the pair breaking [110] edge. These non-zero values of the phase of the $d$-wave pairing amplitude near the edge imply that this phase cannot be gauged away and, consequently, time-reversal symmetry is broken \cite{Sigrist91}. Additionally, the modulating nature of this phase shows that the translational symmetry along the [110] edge is also broken. These modulations define a phase crystal state \cite{Holmvall20}, previously found in the absence of strong correlations in both quasi-classical theory \cite{Vorontsov09,Hakansson2015} and BdG \cite{Wennerdal20}. Clearly, the phase crystal also thrives in the presence of the strong correlations found in the cuprates, and remarkably here also features multiple new distinct properties, as we report below. Before proceeding we also note that the modulating phase of the superconducting order additionally results in circulating orbital currents which we demonstrate in the Supplementary Material (SM).

To start characterizing these new properties, we investigate in Fig.~\ref{fig:cleanPC}(c,d) the behavior of the relative change of the magnitude of the $d$-wave pairing amplitude, $\delta\left|\Delta_{d}\right|$, and its phase, $\theta$, along $x_{\perp}$, i.e.~perpendicular to the pair breaking edge. 
As seen, the length scale $\lambda_{\perp}$, at which the phase $\theta$ decays to its zero bulk value, is the same as the healing length of $\left|\Delta_{d}\right|$, with and without strong correlations. But notably, $\lambda_{\perp}$ is much shorter when we appropriately include the strong correlations. The reason for this dramatically shorter $\lambda_{\perp}$ is two-fold: 
First, strong correlations renormalize the hopping amplitudes $t_{ij}$ and the super-exchange interaction $J$ through Gutzwiller factors in the bulk. Hence the bulk superconducting coherence length is reduced, which automatically leads to a decreasing healing length of $\left|\Delta_{d}\right|$. We however note that these factors are not significantly different between the bulk and the edge.
Secondly, strong electronic repulsion suppresses the charge fluctuations formed at the edge due an increased proximity to the Mott insulating normal state, as previously also established for local impurities \cite{Tang15}. This suppression of the charge fluctuation is mimicked by the local Hartree shift (see Methods), which is notably different on the edge compared to the bulk. As a result, the charge density heals to its bulk value over a short distance and also the healing length of $\left|\Delta_{d}\right|$ is closely tied to the healing length of the charge density in a strongly correlated state \cite{Tang15,Chakraborty14}. 
In fact, even near the edge we find that $\left|\Delta_{d}\right|$ is closely following the relative increase $\delta \rho$ of the charge density, as seen in Fig.~\ref{fig:cleanPC}(c). 
To summarize, both a short bulk superconducting coherence length and a short charge density healing length result into a short healing length of $\left|\Delta_{d}\right|$ into the bulk, and consequently a short $\lambda_{\perp}$. 
In contrast, $\lambda_{\perp}$ in BdG \cite{Wennerdal20} (and also quasiclassical results \cite{Holmvall20}) is only related to the bulk superconducting coherence length, with no relation to the charge density fluctuations, as also clearly seen in Fig.~\ref{fig:cleanPC}(d). The strikingly short $\lambda_{\perp}$ with strong correlations has important consequences for the phase crystal as its modulations along the edge is set by an intricate energy balance: A spatially varying $\theta$ in the bulk costs kinetic energy, whereas such variations in $\theta$ along the edge instead lower the free energy through a Doppler shift of the zero-energy states \cite{Holmvall18,Holmvall20}. Hence, the very short $\lambda_{\perp}$ in GIMT automatically gives a very short modulation wavelength of the phase crystal along the edge, as is evident in Fig.~\ref{fig:cleanPC}(a).

Another striking feature of strong correlations is the appearance of an extended $s$-wave pairing amplitude $\left|\Delta_{s}\right|$ at the edge with a modulation length scale set by that of the phase crystal, as seen in Fig.~\ref{fig:cleanPC}(e), while the $s$-wave amplitude is always negligible in the bulk. Although not shown in Fig.~\ref{fig:cleanPC}(e), the phase of the $s$-wave pairing amplitude is nearly constant in all regions where its magnitude is finite, with a value of $\pi/2$ relative to the phase of the $d$-wave amplitude. We attribute the existence of this $s$-wave component to the very short modulation wavelength of the phase crystal as it gives a large number of phase nodes. These nodes lead to some zero-energy states still persisting in the phase crystal, although their degeneracy is reduced. At low enough temperatures, these remaining zero-energy states make the system thermodynamically unstable, with the consequence that the extended $s$-wave state is developed, gapping out the last states and generating a full energy gap. Ignoring strong correlations, the phase crystal modulation wavelength is instead longer, no extended $s$-wave pairing amplitude emerges, and consequently the energy spectrum is never fully gapped \cite{Hakansson2015,Wennerdal20}. Here we note that the time-reversal symmetry breaking due to the formation of the phase crystal state can technically also generate subdominant odd parity pairing amplitudes, as in superconductor-magnet hybrid systems \cite{Nakosai13,Chen15}. However, with the interaction strength $J$ limited to the spin-singlet channel in the `t-J' model and the cuprate superconductors, such pairing amplitudes never appear in the Hamiltonian and thus cannot affect the energetics\cite{Baskaran2002,Schmidt2018}.
Similarly, no longer range spin-singlet even parity pairing amplitudes, beyond the considered nearest neighbor $d$- and $s$-wave pairing amplitudes, are important as they do not either appear in the Hamiltonian in Eq.~\eqref{eq:tJ}.

\begin{figure}[t]
\includegraphics[width=1.0\linewidth]{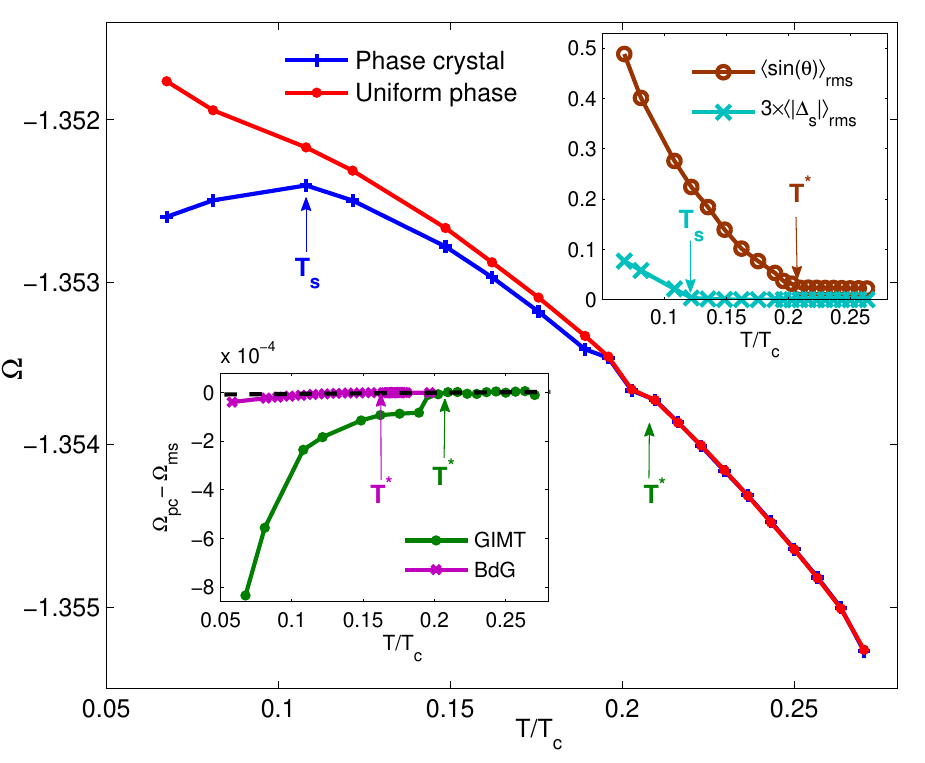} \caption{{\bf {Thermodynamics for clean sample.}} Free energy $\Omega$ calculated in GIMT for phase crystal and uniform phase states as a function of scaled temperature $T/T_c$, where $T_c$ is the $d$-wave superconducting transition temperature. Green arrow indicates the phase crystal transition temperature $T^*$. Upper inset shows the phase transition temperatures $T^*$ and $T_s$ obtained by plotting the phase $\langle \textrm{sin}(\theta) \rangle_{\textrm{rms}}$ and the magnitude of the extended $s$-wave pairing amplitude $\langle \left|\Delta_s\right| \rangle_{\textrm{rms}}$, respectively, averaged along the [110] edge (a very small temperature-independent value of $\Delta_s$ of the uniform phase state has been subtracted to extract $T_s$). Lower inset shows difference in free energy between the phase crystal state, $\Omega_{\rm pc}$, and the metastable uniform phase state, $\Omega_{\rm ms}$, within GIMT and BdG. The energy jump in GIMT near $T=T^*$ is mainly from the internal energy $E_{\rm g}$.}
\label{fig:thermoplot} 
\end{figure}

\begin{figure}[h]
\includegraphics[width=1.0\linewidth]{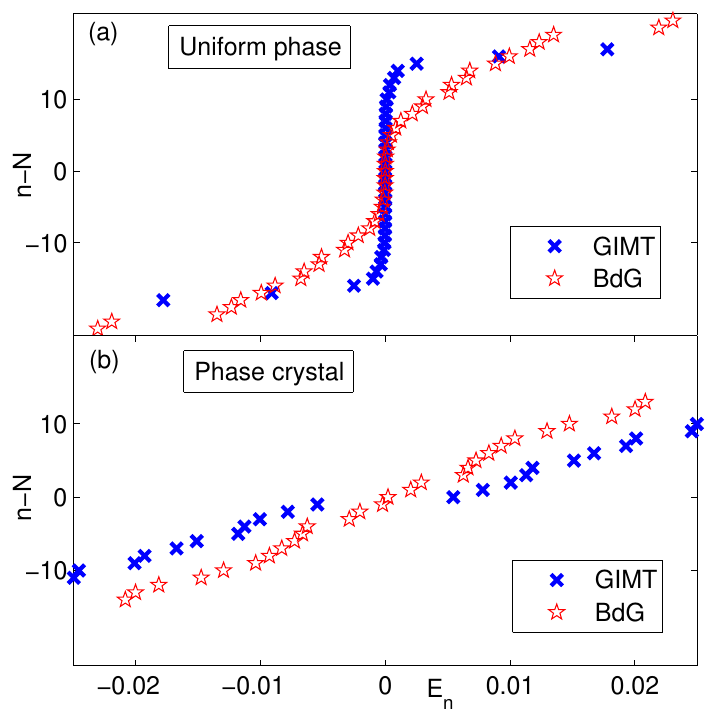} \caption{{\bf {Distribution of low-energy eigenvalues for clean sample.}} Eigenvalues $E_n$ within GIMT and BdG in the uniform phase ({\bf{a}}) and phase crystal ({\bf{b}}) states at a low temperature, $T=0.08T_c<T_s<T^*$. Strong correlations generate a higher number of zero-energy states in the uniform phase in ({\bf{a}}). The phase crystal state shifts most zero-energy states to finite energies within both BdG and GIMT, but GIMT displays larger shifts and a full energy gap ($\textrm{min}\{E_n\}=0.006$) due to the formation of co-existing $s$-wave state. Here, the eigenindex $n$ is shifted by the total number of lattice sites $N$.
}
\label{fig:thermoplot2} 
\end{figure}

\begin{figure*}[t]
\includegraphics[width=1.0\linewidth]{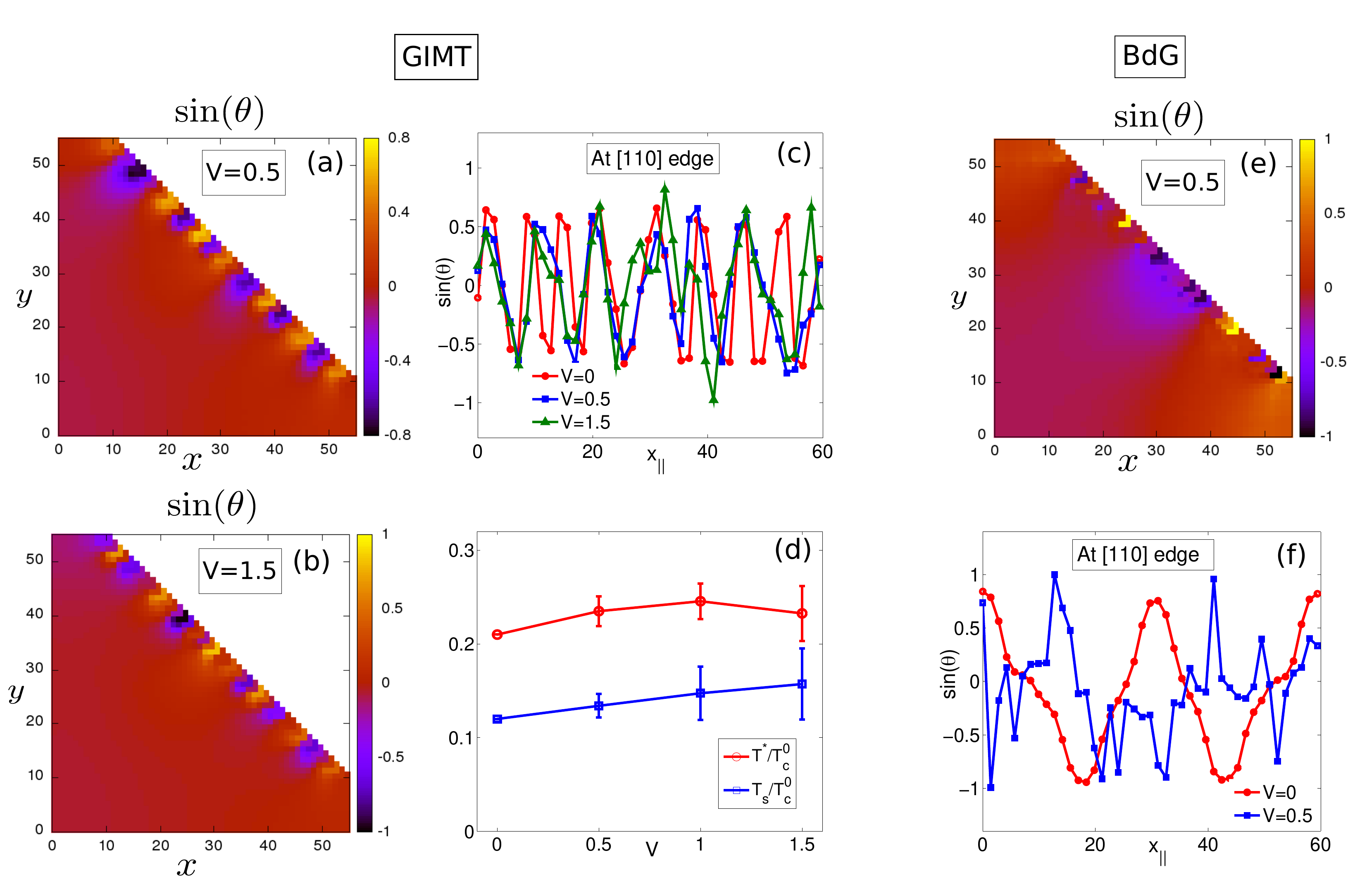} \caption{{\bf {Phase crystals for disordered sample.}} Color density maps of $\sin(\theta)$ in both GIMT ({\bf{a}},{\bf{b}}) and BdG ({\bf{e}}) for $T=0.32T^*<T_s$. Line plots of the phase modulations along the pair breaking [110] edge as a function of $x_{\parallel}$ are shown for GIMT ({\bf{c}}) and BdG ({\bf{f}}) for different disorder strengths $V$. 
The phase crystal is stable even in the presence of strong disorder within GIMT but does not survive even moderately weak disorder if strong correlation effects are not included. 
Phase crystal transition temperature $T^*$ and extended $s$-wave transition temperature $T_s$ are shown in ({\bf{d}}), averaged over five disorder configurations, both scaled by the $V=0$ superconducting transition temperature $T_c^{0}$. Error bars indicate the standard deviation about the mean disorder averaged value.}
\label{fig:disorderpc} 
\end{figure*}

Having found a phase crystal state with accompanied $s$-wave pairing at low temperature in the cuprate superconductors, we next investigate how this exotic state develops with temperature and, in particular, how it compares with a uniform state (say $\theta =0$) throughout the superconductor, as that is also a viable superconducting solution to Eq.~\eqref{eq:tJ}.
In order to investigate the thermodynamic properties, we calculate the free energy $\Omega=E_{\rm g}-TS$, where $E_{\rm g}$ is the internal energy corresponding of the Hamiltonian Eq.~\eqref{eq:tJ} and $S$ is the entropy (see Methods). In Fig.~\ref{fig:thermoplot} we plot $\Omega$ for both the phase crystal state and the uniform phase state. 
The phase crystal state has a lower free energy than the uniform phase state for all temperatures below $T^*$. Thus, $T^*$ defines the transition temperature of the phase crystal state, with its spontaneous breaking of both time-reversal and translational symmetry. At an even lower $T=T_s$, the phase crystal free energy shows a marked downturn. This temperature corresponds to the appearance of the modulating $s$-wave pairing inside the phase crystal state, see upper inset in Fig.~\ref{fig:thermoplot}.
Finally, to understand the effects of strong correlations on the stability of the phase crystal state, we show the free energy difference between the phase crystal state and the (metastable) uniform phase state in the lower inset of Fig.~\ref{fig:thermoplot} for both GIMT and BdG calculations. The energy gain due to the formation of the phase crystals is much larger in GIMT compared to BdG. We further expect the enhancement of the energy gain in GIMT to be even more pronounced for a lattice with larger pair breaking [110] edges than the one considered here, since phase crystal is only occurring at the pair breaking edge. Also, as indicated by arrows, $T^* {\rm (GIMT)}>T^* {\rm (BdG)}$. Both the large free energy gain and the higher $T^*$ demonstrate that the strong correlations give a notably enhanced thermodynamic stability of the phase crystal state.

The remarkably increased thermodynamic stability of the phase crystal state with strong correlations can be understood by looking at the eigenvalues of the Hamiltonian Eq.~\eqref{eq:tJ}. In Fig.~\ref{fig:thermoplot2} we plot the distribution of the low-energy eigenvalues $E_n$, where $n$ is the eigenindex, at a low temperature, for both the phase crystal and metastable uniform phase states and with and without strong correlations. 

In the uniform-phase state, zero-energy Andreev bound states forms at [110] edges \cite{Kashiwaya00,Lofwander01}. The origin of these zero-energy states lies in the change of the sign of the $d$-wave superconducting order parameter experienced by quasiparticles Andreev scattering off the [110] edge \cite{Kashiwaya00}. Such a sign change does not occur at edges parallel to the crystallographic axes, such as the [100] edge, and hence the zero-energy states are only existing at the [110] edge in our sample. Moreover, since scattering trajectories are dependent on the quasiparticle band structure and its superconducting nodes, there exists an inherent band structure effect.
Alternatively, and completely equivalently, the zero-energy states at [110] edges in a $d$-wave superconductor can be seen as having a topological origin set by the nodal structure of the quasiparticle band structure \cite{Ryu02,Sato11}. 
More specifically, the number of zero-energy edge states is equal to the number of states in the region of the edge Brillouin zone bounded by the projections of the superconducting nodes in the bulk band structure \cite{Potter14}.
It is here important to emphasize that the topological nature of the edge states thus depends both on the orientation of the edge and the nodal structure band structure, which distinguishes the behavior topological edge states of a $d$-wave superconductor from the edge states of a topological insulator. 
As expected, we find zero-energy states in the uniform phase of both GIMT and BdG, as shown in Fig.~\ref{fig:thermoplot2}(a). However, the number of zero-energy states is significantly increased with strong correlations in GIMT.  

We can relate this increase of the number of zero-energy states to a notable increase in the charge density at the edge always found when including strong correlations within GIMT, but not in BdG. This increase of the charge density at the edge can in turn be understood using a simple argument: Strong electronic repulsion tries to keep the electrons as far apart from each other as possible, or in other words, strong correlations smear out any charge accumulation. However, at the edge, this smearing effect is always reduced due to a reduced number of neighbors, and, as a result, the electrons accumulate at the edge compared to the bulk. In a sense, the enhancement of charge density at the edge can thus be thought of as a `surface tension' caused by strong correlations. Notably, this result asks for a reconsideration of the straightforward use of the bulk-boundary correspondence in nodal superconductors, which is used to build a topological relation between the gap nodes and the number of zero-energy edge states \cite{Sato11,Schnyder15}. A straightforward use of the bulk-boundary correspondence for the [110] edge of a $d$-wave superconductor would mean that the number of zero-energy edge states is set by the projections of the gap nodes in the bulk band structure onto the edge Brillouin zone \cite{Potter14}. However, the change in the edge charge density when including strong correlations results in a different effective band structure near the edge compared to that of the bulk. As a consequence, a reasonable understanding of the number zero-energy edge states can only be established if bands corresponding to the edge charge density are considered, instead of those corresponding to the bulk charge density, see SM for additional details
Thus, the bulk-boundary correspondence should be applied using the modified edge band structure as the new ``bulk" part in the correspondence. This is a clear example where the topological relation between the superconducting gap nodes and the zero-energy edge states in nodal superconductors needs to be reformulated in the presence of strong correlations. 

In the phase crystal state, eigenvalues are also shifted to finite energies, see Fig.~\ref{fig:thermoplot2}(b), making it the ground state below $T^*$ in both GIMT and BdG. Still, there exist three crucial effects of strong correlations. First, the number of zero-energy states is significantly larger in the GIMT uniform phase than in the BdG uniform phase and consequently the energy gain in the phase crystal state, due to the shift of these states to finite energies, is much larger in GIMT. Secondly, due to the presence of the extended $s$-wave order, the GIMT phase crystal state features a full energy gap below $T_s$, while there are always zero-energy states present in BdG calculations. Finally, even outside the full energy gap, the eigenvalues of the negative energy states are found at lower energies in GIMT compared to in BdG, and also display a moderate temperature dependence, see SM for additional data. These effects all contribute to making the phase crystal state more heavily preferred in the presence of strong correlations. We thus find that strong electronic correlations are crucial to generate both the correct stability and the full energy gap of the phase crystal state.

{\noindent {\textbf {Dirty superconductor edge.}}} After having established a fully gapped phase crystal state at pair breaking edges of clean $d$-wave superconductors when accounting for strong correlations, we turn to another important aspect of cuprate superconductors, that of disorder. The bulk both possesses intrinsic disorder \cite{Howald01} and disorder is generated with chemical substitution and doping \cite{Alloul09}. For systems with edges, disorder is even more important as it inevitably appears while preparing any edge. Since we are here primarily interested in edge properties, we only consider non-magnetic disorder for which the bulk $d$-wave superconducting state has been shown to be robust when including strong correlation effects \cite{Garg08,Chakraborty14,Tang16}.

\begin{figure}[t]
\includegraphics[width=1.0\linewidth]{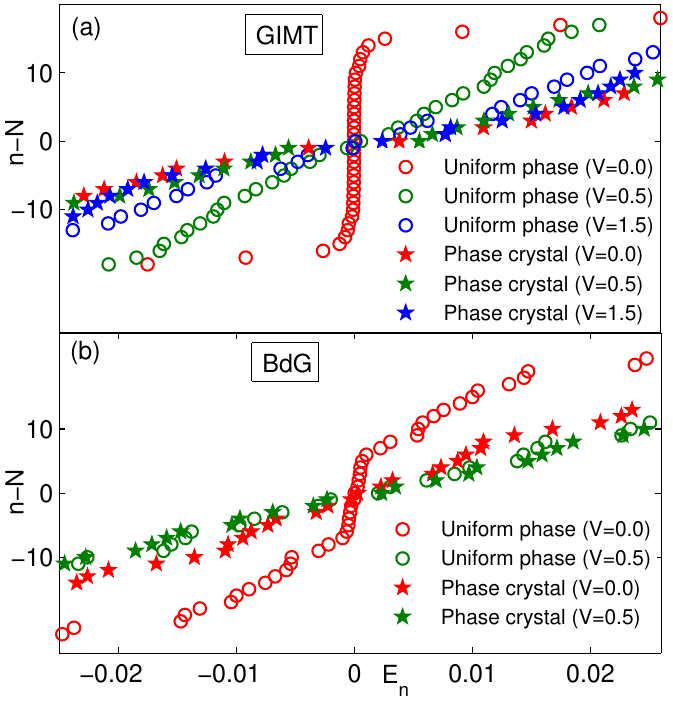} \caption{{\bf {Distribution of low-energy eigenvalues for disordered sample.}}  Eigenvalues $E_n$ within GIMT ({\bf{a}}) and BdG ({\bf{b}}) at the same temperature as in Fig.~\ref{fig:disorderpc} ($T=0.32T^*<T_s$), with stars (circles) for the phase crystal (uniform phase) state at different disorder strengths $V$. The phase crystal state within GIMT in ({\bf{a}}) is stable for all $V$, while the eigenvalues for the phase crystal and uniform phase states almost overlap within BdG in ({\bf{b}}), corroborating the fact that the phase crystal does not survive disorder when strong correlations are ignored.
}
\label{fig:diseig} 
\end{figure}

In Fig.~\ref{fig:disorderpc}(a,b) we show the evolution of the phase of the $d$-wave pairing amplitude at a temperature within the phase crystal state for different disorder strengths $V$. The phase modulations near the pair breaking edge persist in the whole range from weak to strong disorder strength $V = 1.5$. This is also seen in the line plots of the phase at the pair breaking edge in Fig.~\ref{fig:disorderpc}(c). Moreover, the disorder averaged values of the phase crystal transition temperature $T^*$ and the accompanying extended $s$-wave transition temperature $T_s$ also remain constant with disorder, see Fig.~\ref{fig:disorderpc}(d). Thus, the phase crystal state is remarkably robust to disorder when strong correlations are appropriately included. The almost complete disorder insensitivity of the phase crystal is very surprising. It is true that strong correlations have previously been seen to be the reason for the robustness of $d$-wave superconductors to disorder \cite{Garg08,Chakraborty14,Tang16}, but this robustness has always involved the magnitude of the pairing amplitude. In contrast, the phase crystal state exists only near pair breaking edges where the magnitude is suppressed and it is here instead the superconducting phase that is completely robust to disorder.

To further imprint the importance of strong correlations for the disorder robustness, we plot in Fig.~\ref{fig:disorderpc}(e,f) the phase of the $d$-wave pairing amplitude obtained within BdG for $V=0.5$. Even for this relatively weak $V=0.5$, where bulk $d$-wave superconductivity clearly still thrives \cite{Chakraborty14}, the edge phase modulations are almost completely disrupted. Thus, in the absence of strong correlations, the phase crystal state is clearly extremely sensitive to disorder, much more so than the bulk $d$-wave state. This further emphasizes the remarkable role of strong correlations in the stability of the phase crystal state. We have verified that these results hold  also for other models of disorder, see SM.

The disorder robustness of the phase crystal state is even more surprising if we consider the energetic origin of the phase crystal state: a large degeneracy of zero-energy states in the uniform phase state. It has recently been shown using both T-matrix calculations \cite{Kalenkov04} and topological arguments \cite{Ikegaya17} that disorder generally weakens zero-energy state degeneracies. This we also see in the uniform phase state in Fig.~\ref{fig:diseig}(a), where we plot the low-energy eigenvalues $E_n$. In the uniform state, the spectrum of eigenvalues is changed by disorder and the number of zero-energy states is indeed rather dramatically reduced.
Based on this, we would naively expect that the phase crystal state would quickly disappear with increasing disorder. Instead, we see that disorder has very little effect on the phase crystal states and its low-energy states, when we appropriately include strong correlation effects in Fig.~\ref{fig:diseig}(a). As a consequence, the energy gain due to the formation of the phase crystal state is still higher than the disorder-induced shift of the energy spectrum in the uniform phase state. This results in the phase crystal state being robust to disorder in the presence of strong correlations, even though the degeneracy of the zero-energy states in the uniform state is nearly eliminated.
In contrast to the disorder robustness when strong correlations are appropriately included, we find in Fig.~\ref{fig:diseig}(b) that the BdG eigenspectra of the uniform and phase crystal states start to overlap even for small disorder. As a result, the formation of a phase crystal state does not reduce the total free energy and thus the phase crystal state is destroyed even at weak disorder when strong correlations are ignored.

The striking disorder robustness of the phase crystal state in the presence of strong correlations has two underlying reasons. First, the enhanced number of zero-energy states in the uniform phase state of GIMT results in a larger energy gain of the phase crystal state for the clean sample, compared to the non-correlated case. This energy gain is clear from the lower inset of Fig.~\ref{fig:thermoplot}. All by itself, this large energy gain already suggests that the phase crystal within GIMT can withstand disorder up to a larger disorder strength compared to BdG. Second, strong electronic repulsion smears out the charge inhomogeneities caused by the presence of disorder. This results in a strongly weakened effective disorder \cite{Chakraborty14}. These two features cumulatively make the phase crystal state robust to disorder in the presence of strong correlations.

Our results demonstrate that the combined effects of strong correlations, topology, and disorder can be very non-intuitive. The phase crystal state, which initially is formed due to a large degeneracy of topologically protected zero-energy states, is extremely robust to disorder but only in the presence of strong correlations. This result holds even though the original degeneracy of the zero-energy states is drastically reduced by disorder.  

\begin{figure}[t]
\includegraphics[width=1.0\linewidth]{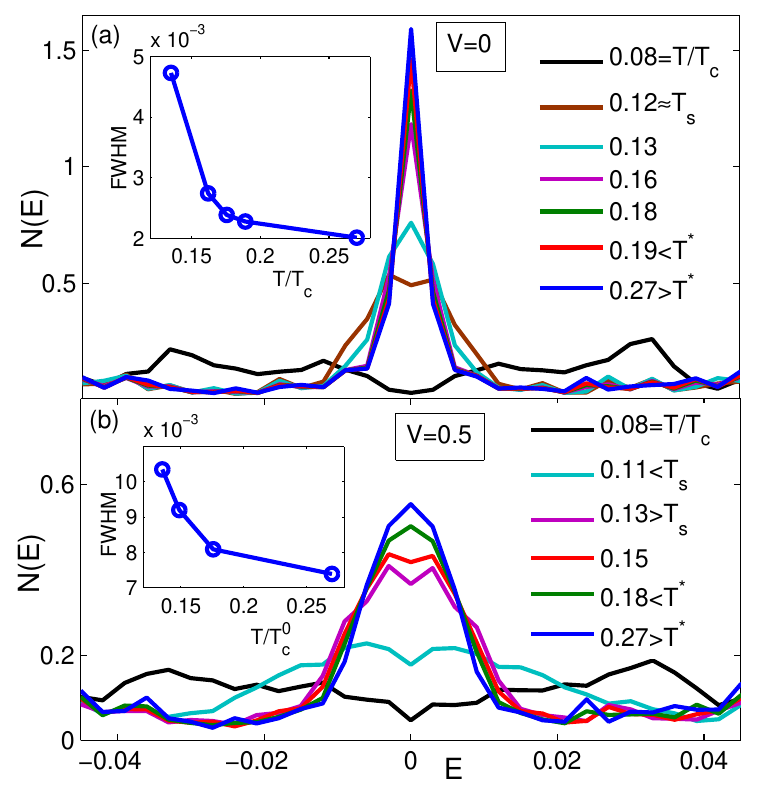} \caption{{\bf {Low energy density of states.}} Spatially averaged density of states $N(E)$ within GIMT at different temperatures for clean (a) and disordered samples (b). 
$N(E)$ is extracted with a constant broadening parameter, which excludes the temperature broadening caused by the Fermi-Dirac distribution.
A clear zero-bias peak is seen for $T>T_s$, with insets showing its full width at half maximum, FWHM. The increase of FWHM with decreasing temperature shows that the zero-bias peak is broadened beyond the Fermi-Dirac distribution effects, i.e.~a temperature-independent effect.
For $T<T_s$, $N(E)$ develops a full gap (small remnant $N(E)$ inside the gap is due to the broadening of the delta-function). Plots for finite disorder are averaged over five disorder configurations.
}
\label{fig:dos} 
\end{figure}

\noindent {\textbf {Comparison with experiments.}} Above we established the existence of a disorder robust phase crystal state with a full gap at the lowest temperatures at edges of strongly correlated high-temperature cuprate superconductors.
These findings have direct experimental application in superconducting cuprate devices since the presence of zero-energy edge states without the phase crystal state should, if indeed present, significantly affect transport properties.
In fact, different experiments on cuprate devices have for a long time given contradictory results, with no previous viable theoretical explanation.
At intermediate temperatures, tunneling experiments across cuprate-normal metal junctions unanimously see a large zero-bias conductance peak \cite{Wei98,Kashiwaya00}, consistent with the existence of many zero-energy edge states. However, at lower temperatures, some experiments reveal a full gap in the tunneling spectra \cite{Covington97,Dagan01,Gustafsson13}, whereas others only see a so-called temperature-independent broadening \cite{Alff97,Neils02}, i.e.~broadening beyond standard thermal broadening through the Fermi-Dirac distribution, of the zero-bias conductance peak. This dichotomy at low temperatures has previously been notoriously hard to explain theoretically.
A set of theories, based on either spontaneous formation of subdominant $s$-wave superconducting order, \cite{Black-Schaffer13,Nagai17} magnetic order \cite{Honerkamp00,Potter14}, or supercurrents \cite{Lofwander00} all give only a split in the zero-energy density of states at low temperatures. 
On the other hand, a phase crystal state in the absence of strong correlations only explains the temperature-independent broadening of the zero-energy states \cite{Hakansson2015} but not the full gap. By including strong correlation effects we remedy these theory shortcomings and can now explain all experimental data.

To connect to the experiments, in Fig.~\ref{fig:dos} we show the spatially averaged low-energy density of states $N(E)$ obtained within GIMT at different temperatures in clean (a) and disordered (b) samples. To numerically evaluate $N(E)$, we choose a small and temperature independent energy smearing in $N(E)$ (see Methods for details). We do not expect broadening due to any other scattering processes to be temperature dependent for the low temperatures considered in this work, as the energy scales associated with these temperatures are much smaller than the superconducting gap. We find that for both clean and disordered samples, at high temperatures, $T>T^*$, the zero-energy states of the uniform phase state persist, giving a pronounced zero-bias peak in the $N(E)$. 
Below a lower temperature, $T^*$, the phase crystal state develops, which induces a broadening of the lowest energy peak. This is most clearly seen in the temperature evolution of the full width at half maximum (FWHM) of the zero-energy peak, which we plot in the insets of Fig.~\ref{fig:dos}. 
Most importantly, the zero-energy peak broadening increases with decreasing temperature and exists independently of the thermal broadening (which is not included), therefore being a temperature-independent effect. It is caused by an increase in the Doppler shift of the low energy states in the phase crystal state with lowering temperatures, see also SM for the explicit temperature dependence of the eigenvalues.
Finally, at an even lower temperature, $T_s$, an additional extended $s$-wave component appears causing a full energy gap to develop. 
Thus, it is very feasible to experimentally measure, across different devices with their different $T^*$ and $T_s$ due to microscopic details, both a full energy gap and a temperature-independent broadening of the zero-bias conductance peak at low temperatures, while at intermediate temperatures consistently see the large zero-bias peak of the uniform state. This illustrates how our findings likely resolve the long-standing experimental dichotomy and provide future guidelines for the design of superconducting devices.

\noindent {\textbf {Discussion.}} The strong disorder robustness of the phase crystal state in strongly correlated superconductors makes it a very promising candidate for finally explaining the physics of boundaries and edges in the cuprate superconductors. The disorder robustness of the phase crystal state is also remarkable from the point of view that it concerns the phase of the superconducting pairing, not the magnitude. Conventional $s$-wave superconductors have for a long time been known to be highly robust against disorder thanks to the celebrated Anderson's theorem \cite{Anderson59}. In the presence of strong correlations, an analogy of the Anderson's theorem has recently been constructed for $d$-wave superconductors \cite{Ghosal18}. However, these results both only concern the magnitude of the superconducting pairing.
The phase crystal state on the other hand is characterized by its strong modulation of the phase of the superconducting pairing and it even only exists in regions where the magnitude is suppressed. 
Thus, there exists no established reason to expect a phase crystal state to be anything but unstable towards disorder. This disorder sensitivity is even verified in the absence of strong correlations, where the phase crystal is very easily destroyed by disorder. It is only when including strong correlation effects that the phase crystal state is stabilized to the degree that it can be present in a real material. Thus our results establish the importance of strong correlations for disorder robustness of superconducting phase modulations.
Since disorder robustness is often used to predict the underlying pairing symmetry of various superconductors, our findings can additionally play a pivotal role in determining the pairing symmetry of newly discovered superconductors such as nickelates or twisted bilayer graphene, where electronic correlations are considered to be strong \cite{Zhang20,Cao18a}.

~

\footnotesize
\small
\noindent {\textbf {Methods}} 

\noindent In this section, we first provide details of the two methods, GIMT and BdG, implemented to solve the Hamiltonian in Eq.~\eqref{eq:tJdis} to extract the self-consistent superconducting pairing amplitude, eigenvalues, and thermodynamic quantities. Then we provide values and motivations for the lattice geometry and different parameters used to obtain the results. 

{\noindent {\textbf {GIMT.}}} Solving the (disordered) `t-J' Hamiltonian in Eq.~\eqref{eq:tJdis} requires handling the projection operators that prohibit formation of double occupancy on any lattice site. Within the Gutzwiller inhomogeneous mean-field theory (GIMT) the effects of this projection is treated locally, i.e.~at each site, using the Gutzwiller approximation. To implement the Gutzwiller approximation, we first consider the Gutzwiller wave function\cite{Gutzwiller63,Paramekanti01,Anderson04} $|\psi\rangle = {\cal P}|\psi_0\rangle$, where $|\psi_0\rangle$ is the ground state wave function in the Hilbert space that allows double-occupancy and ${\cal P}$ is the projection operator: ${\cal P}=\prod_i{\cal P}_i$, with ${\cal P}_i=\gamma^{{n_i}/2}_i(1-n_{i \uparrow} n_{i \downarrow})$, where $\gamma_i$ are the local fugacity factors \cite{Fukushima08} obtained by demanding conservation of the local electron densities. Expectation values of a general operator $O$ are likewise ${\langle O \rangle}_0 =  {\langle\psi_0|O|\psi_0\rangle}/{\langle\psi_0|\psi_0\rangle}$ and ${\langle O \rangle} =  {\langle\psi|O|\psi\rangle}/{\langle\psi|\psi\rangle}$ in the unprojected and projected Hilbert space, respectively. Within the Gutzwiller approximation, expectation values of the different terms of the Hamiltonian Eq.~\eqref{eq:tJdis} are assumed to be related by \cite{Zhang88,Edegger07,Ko07}
\begin{eqnarray}
\langle c^{\dagger}_{i \sigma} c_{j \sigma} \rangle &\approx& g^t_{ij}\langle c^{\dagger}_{i \sigma} c_{j \sigma} \rangle_0, \nonumber \\
\langle \mathbf{S}_i \cdot \mathbf{S}_j \rangle \approx g^{J}_{ij}\langle \mathbf{S}_i \cdot \mathbf{S}_j \rangle_0,&&\langle n_i n_j \rangle \approx \langle n_i n_j \rangle_0,
\label{eq:gut1}
\end{eqnarray}
where $g^t_{ij}$ and $g^{J}_{ij}$ are known as the Gutzwiller renormalization factors. In the absence of magnetic order and ignoring intersite correlations, $g^t_{ij}$ and $g^{J}_{ij}$ are given in terms of the local hole doping $x_i$ \cite{Garg08,Fukushima08}:
\begin{equation}
g^t_{ij}=\sqrt{\frac{4x_ix_j}{(1+x_i)(1+x_j)}} ~~{\rm and}~~ g^{J}_{ij}=\frac{4}{(1+x_i)(1+x_j)}.
\label{eq:gut2}
\end{equation}
Using the relations obtained by the Gutzwiller approximation in Eq.~\eqref{eq:gut1}, the internal energy of the Hamiltonian in Eq.~\eqref{eq:tJdis}, $E_{\rm g}=\langle\psi_0|{\cal H}|\psi_0\rangle$, can be expressed as
\begin{eqnarray}
&&E_{\rm g}=\sum_{ij} 2t_{ij} g^t_{ij} \left( \tau_{ij}+\tau_{ij}^{*} \right) + \sum_{i}\left( V_i-\mu \right) \rho_i \nonumber \\
&&-\frac{J}{2} \sum_{\langle ij \rangle} \left\{ \left( 3g^{J}_{ij} - 1 \right) \left|\tau_{ij}\right|^2 + \frac{\rho_i\rho_j}{2}+\frac{\left( 3g^{J}_{ij} + 1 \right)}{4}\left|\Delta_{ij}\right|^2\right\},\nonumber \\
\label{eq:energy}
\end{eqnarray}
where $\rho_i=\sum_{\sigma} \langle n_{i \sigma} \rangle = \sum_{\sigma} \langle n_{i \sigma} \rangle_0$ is the local electron density, $\Delta_{ij} = \langle c_{j \downarrow} c_{i \uparrow} \rangle_0 - \langle c_{j \uparrow} c_{i \downarrow} \rangle_0$ is the spin-singlet Cooper pairing amplitude on each nearest neighbor bond, and $\tau_{ij} = \langle c_{i \downarrow}^{\dagger} c_{j \downarrow} \rangle_0 = \langle c_{i \uparrow}^{\dagger} c_{j \uparrow} \rangle_0$ is the particle-hole bond expectation value. The local doping is given by $x_i=1-\rho_i$. In writing Eq.~\eqref{eq:energy}, we assume full spin rotational symmetry (due to no magnetic order) and no spin-triplet superconducting order, both appropriate for cuprate superconductors. Finally, the extremum of $E_{\rm g}$ with respect to the different mean-field variables gives the GIMT Hamiltonian \cite{Wang06,Christensen11,Fukushima09,Yang09,Chakraborty14}
\begin{eqnarray}
&&{\cal H}_{\rm GIMT}=\sum_{ij \sigma} \frac{\partial E_{\rm g}}{\partial \tau_{ij}} c^{\dagger}_{i \sigma} c_{j \sigma} + \textrm{H.c.}  \nonumber \\ 
&+& \sum _{ij} \frac{\partial E_{\rm g}}{\partial \Delta_{ij}} \left(c^{\dagger}_{i \uparrow} c^{\dagger}_{j \downarrow}-c^{\dagger}_{i \downarrow} c^{\dagger}_{j \uparrow}\right) + \textrm{H.c.} + \sum_{i} \frac{\partial E_{\rm g}}{\partial \rho_{i}} n_{i},
\label{eq:meanfield1}
\end{eqnarray}
which, when combining Eqs.~\eqref{eq:energy} and \eqref{eq:meanfield1}, can be written as
\begin{eqnarray}
&&{\cal H}_{\rm GIMT}=\sum_{i,\delta,\sigma} \left\{ -t g^t_{i,i+\delta}-W^{\rm FS}_{i\delta} \right\} c^{\dagger}_{i \sigma} c_{i+\delta \sigma} \nonumber \\
&&+\sum_{i,\tilde{\delta},\sigma} t^{\prime} g^t_{i,i+\tilde{\delta}} c^{\dagger}_{i \sigma} c_{i+\tilde{\delta} \sigma} + \sum_{i} \left(V_i-\mu+\mu_i^{\rm HS}\right) n_{i}\nonumber \\
&&-\sum_{i,\delta}\left\{\frac{ J\left(3g^{J}_{i,i+\delta} + 1\right)}{16} \Delta_i^{\delta}\left(c^{\dagger}_{i\uparrow} c^{\dagger}_{i+\delta \downarrow}+c^{\dagger}_{i+\delta \uparrow} c^{\dagger}_{i \downarrow}\right)+\textrm{H.c.}\right\},
\label{eq:meanfield2}
\end{eqnarray}
where now nearest neighbor sites of $i$ are denoted as $j=i+\delta$, $\delta=\pm x, \pm y$, and the next-nearest neighbor sites as $j=i+\tilde{\delta}$, $\tilde{\delta}=\pm (x \pm y)$. Consequently, $\Delta_{ij}$, which resides on nearest neighbor bonds, is denoted as $\Delta_i^{\delta}$, and similarly for $\tau$. Here, $W^{\rm FS}_{i\delta}$ and $\mu_i^{\rm HS}$ are the Fock and Hartree shifts, respectively,  given by
\begin{eqnarray}
W_{i \delta}^{\rm FS}&=&\frac{J}{4} \left(3g_{i,i+\delta}^{J}-1\right) \tau_{i}^{\delta}, \nonumber \\
\mu_i^{\rm HS}&=&-2t\sum_{\delta} \frac{\partial{g_{i,i+\delta}^{t}}}{\partial{\rho_i}} \left( \tau_i^{\delta}+(\tau_i^{\delta})^*\right) \nonumber \\
&+&2t'\sum_{\tilde{\delta}} \frac{\partial{g_{i,i+\tilde{\delta}}^{t}}}{\partial{\rho_i}} \left( \tau_i^{\tilde{\delta}}+(\tau_i^{\tilde{\delta}})^*\right) \nonumber \\
&-&\frac{3J}{2}\sum_{\delta} \frac{\partial{g_{i,i+\delta}^{J}}}{\partial{\rho_i}}\left( \frac{1}{4} \left|\Delta_i^{\delta}\right|^2+\left|\tau_i^{\delta}\right|^2 \right) \nonumber \\
&-&\frac{J}{4} \sum_{\delta} \rho_{i+\delta}.
\label{eq:hFshifts}
\end{eqnarray}
The derivatives of the Gutzwiller factors in Eq.~\eqref{eq:hFshifts} are calculated analytically using the expressions in Eq.~\eqref{eq:gut2}. Notably, these derivatives suppress any change in the charge density between the site $i$ and its neighbors $j$ and are thus responsible for reducing charge accumulation within GIMT.

The mean-field Hamiltonian ${\cal H}_{\rm GIMT}$ in Eq.~\eqref{eq:meanfield2} is diagonalized using the Bogoliubov-de Gennes transformations \cite{Zhubook}, $c_{i \sigma}=\sum_{n} ( \gamma_{n \sigma} u_{i,n}-\sigma \gamma^{\dagger}_{n \bar{\sigma}} v_{i,n}^{*})$, where $\gamma^{\dagger}_{n\sigma}$ and $\gamma_{n\sigma}$ are the creation and annihilation operators of the Bogoliubov quasiparticles and $u_{i,n}$ and $v_{i,n}^{*}$ the eigenfunctions with eigenvalues $E_n$,. The resulting eigen-system is then solved self-consistently for all independent local variables: $\Delta_i^{+x}$, $\Delta_i^{+y}$, $\tau_i^{+x}$, $\tau_i^{+y}$, $\tau_i^{y+x}$, $\tau_i^{y-x}$ and $\rho_i$. Finally, in order to analyze the pairing amplitude we project $\Delta_i^\delta$ on its local $d$- and $s$-wave components $\Delta_{d,s}(i)$ as outlined in the main text. In this work we always report on the superconducting pairing amplitudes $\Delta_i^\delta$, as is commonly done in GIMT calculations \cite{Garg08,Christensen11,Yang09,Chakraborty14}, while technically the superconducting order parameter is given by $g^t_{i,i+\delta}\Delta_i^\delta$ \cite{Zhang88,Chakraborty17a}.

With both $\Delta_i^\delta$ and $\tau_i^\delta$ possibly being complex-valued, we obtain self-consistency on both the real and the imaginary parts. A crucial aspect in this process is the initial guess of the self-consistent variables. With real-valued inputs for $\Delta_i^\delta$ and $\tau_i^\delta$, the solution always converges to a real $\Delta_i^\delta$. Thus, a purely real $\Delta_i^\delta$ is always a solution of the Hamiltonian, albeit it might not be the ground state. In order to converge to the true ground state, that might also include complex values of $\Delta_i^\delta$, complex-valued initial guesses of either $\Delta_i^\delta$ or $\tau_i^\delta$ are essential. We have tried many different inputs as starting guesses. All results for the phase crystal state reported in this work are obtained using completely random complex initial inputs for both the amplitude and phase of $\Delta_i^\delta$, while the uniform phase state is obtained using completely random real-valued amplitude inputs with zero phase.

We also extract several different physical properties. In order to compare the thermodynamic stability of different solutions and determine the ground state, we calculate the free energy $\Omega=E_{\rm g}-TS$, where $E_{\rm g}$ is given by Eq.~\eqref{eq:energy} and $S$ is the entropy given by $S=-2\sum_{n}\left[ f(E_n)\ln~f(E_n)+(1-f(E_n))\ln~(1-f(E_n))\right]$, where $f$ is the Fermi-Dirac distribution function. We also calculate the spatially averaged density of states $N(E)=1/N\sum_{i,n}g^{t}_{i,i}(|u_{i,n}|^2\delta(E-E_n)+|v_{i,n}|^2\delta(E+E_n))$. Here $N$ is the total number of lattice sites. To numerically evaluate $N(E)$, we use a Lorentzian with fixed width 0.0015 to calculate the delta-function and implement a small and temperature independent energy smearing. The fixed width explicitly excludes the standard thermal broadening effect caused by the Fermi-Dirac distribution of electrons, allowing us to isolate effects beyond the standard thermal broadening.

{\noindent {\textbf {BdG.}}} To identify the effects of strong electronic correlations, we compare the results of GIMT with a standard Bogoliubov-de Gennes (BdG) calculation, \cite{Zhubook} where the sole effect of the correlations becomes the creation of the superconducting pairing amplitude. Within the BdG method, the (disordered) `t-J' model Eq.~\eqref{eq:tJdis} is thus solved by ignoring the effects of projection, i.e., the Hilbert space does not have any restriction on the formation of double occupancy, and mean-field decoupling of the super-exchange interaction term is only performed in the Cooper pairing channel. As a consequence, the BdG method does not capture the no-double occupancy constraint imposed by the strong correlations, nor does it include the Hartree ($\mu_i^{\rm HS}$) and Fock ($W_{i \delta}^{\rm FS}$) shifts also given by the strong correlations. However, the lattice scale effects (short wavelength fluctuations) that are relevant for high-temperature superconductors with short superconducting coherence lengths are still included in a BdG solution. The resulting BdG mean-field Hamiltonian is given by 
\begin{eqnarray}
&&{\cal H}_{\rm BdG}=\sum_{i,\delta,\sigma} -t c^{\dagger}_{i \sigma} c_{i+\delta \sigma} +\sum_{i,\tilde{\delta},\sigma} t^{\prime} c^{\dagger}_{i \sigma} c_{i+\tilde{\delta} \sigma} + \sum_{i} \left(V_i-\mu\right) n_{i}\nonumber \\
&&-\sum_{i,\delta}\left\{\frac{ J}{4} \Delta_i^{\delta}\left(c^{\dagger}_{i\uparrow} c^{\dagger}_{i+\delta \downarrow}+c^{\dagger}_{i+\delta \uparrow} c^{\dagger}_{i \downarrow}\right)+\textrm{H.c.}\right\}.
\label{eq:BdGH}
\end{eqnarray}
Practically this BdG Hamiltonian is obtained by setting the Gutzwiller factors $g^t_{ij}$ and $g^J_{ij}$ to unity and ignoring the Hartree ($\mu_i^{\rm HS}$) and Fock ($W_{i \delta}^{\rm FS}$) shifts in the GIMT Hamiltonian in Eq.~\eqref{eq:meanfield2}. Then, we follow the same iterative self-consistency treatment as in GIMT, but here self-consistency is only needed for $\Delta_i^\delta$.

\renewcommand{\figurename}{\footnotesize FIG.}
\renewcommand{\thefigure}{\footnotesize \arabic{figure}}
\begin{figure}[t]
\includegraphics[width=0.8\linewidth]{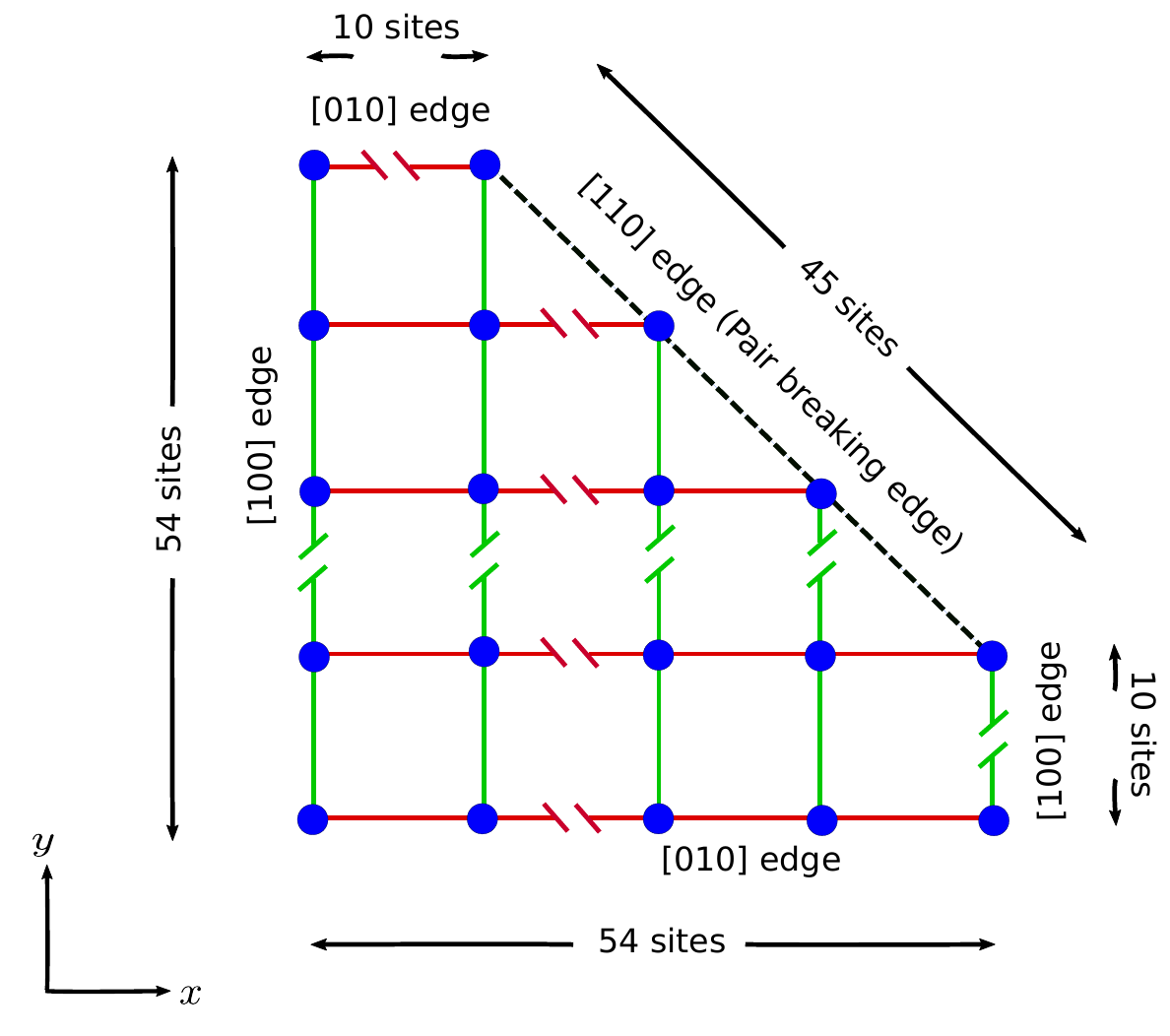}\caption{\footnotesize {\bf{Schematic of lattice geometry.}} Sites (blue) arranged in a square lattice with red and green colors indicating opposite signs of the $d$-wave superconducting pairing amplitude residing on each bond in the bulk. The [110] edge is pair breaking for such superconducting pairing and is shown with a dashed line, while the [010] and [100] edges are non-pair breaking. Kinks indicate that the actual number of lattice sites considered is much larger than what is included in this schematic.}
\label{fig:latticegeo} 
\end{figure}

{\noindent {\textbf {Lattice geometry.}}} We primarily consider the lattice geometry shown in Fig.~\ref{fig:latticegeo}, using $N=1926$ lattice points. This particular lattice geometry gives us an optimum length for the [110] edge (dashed line in Fig.~\ref{fig:latticegeo}) with minimum interference from other edges. This is important since the [110] edge is pair breaking for $d$-wave pairing on nearest neighbor bonds (technically $d_{x^2-y^2}$ symmetry, but here abbreviated as $d$-wave) as the Andreev reflection at this edge mixes opposite signs of the superconducting pairing amplitudes. We have also performed calculations for other system sizes with similar qualitative outcomes, provided the [110] edge is long enough to host multiple modulations of the phase crystal state.

{\noindent {\textbf {Parameters.}}} Within GIMT we set $t=1$ and $t^{\prime}=0.5$ to mimic a normal-state band with Fermi velocities at the anti-nodes being very similar to the nodal ones, which is the case for cuprate materials such as Bi$_2$Sr$_2$CaCu$_2$O$_{8+x}$ (BSCCO) \cite{Norman07}, while only using a minimal number of hopping parameters. We fix the average density to be $\rho_{\rm av}=0.8$ (i.e.~average hole doping 0.2), which gives a doping slightly below the optimal doping for superconductivity for the chosen parameters. Our choice of $\rho_{\rm av}$ allows us to ignore any competing magnetic order (which appears near the extremely underdoped regime) or competing extended $s$-wave superconducting order (which appears in the extremely overdoped regime) in the bulk. Further, we set the super-exchange interaction to $J=0.33$, a value close to the one obtained in scattering experiments on cuprates \cite{Lyons88,Tranquada89}. The chosen value of $J$ corresponds to a large value of $U=12$, mimicking the strong Coulomb interaction in cuprates. In the bulk this gives rise to a bulk superconducting $T_c=0.148$ and $\left|\Delta_{d}\right|=0.236$. We have also checked that the qualitative findings of this work do not change for other lower values of $J$. In terms of the Gutzwiller renormalization factors, we find that they are all significant but that both the hopping amplitudes $t_{ij}$ and the super-exchange interaction $J$ are essentially the same in the bulk as on the edge. However, through the derivatives in Eq.~\eqref{eq:hFshifts}, the local Hartree shift of the chemical potential differs substantially, being around $1.5$ in the bulk but $1.15$ at the edge.

For the BdG calculations we choose the same parameters as used in Ref.~[\onlinecite{Wennerdal20}] to study the phase crystal state. In particular, within BdG we set $t=1$, $t^{\prime}=0.25$, $J=1.4$, and $\mu=0.0$, giving an average density to be $\rho_{\rm av}=1.18$ and the bulk superconducting $T_c=0.122$ and $\left|\Delta_{d}\right|=0.108$. These parameters both reproduce earlier data in the literature for BdG calculations and also generate normal-state band parameters that are very close to those of the GIMT calculations. We note however that both the density and transition temperatures are different compared to in GIMT, but we can still compare the GIMT and BdG solutions by reporting properties rescaled by $T_c$. 
In the SM we provide an alternative BdG (alt-BdG) calculation where we choose the BdG parameters explicitly such that both the bulk average density $\rho_{\rm av}$ and pairing amplitude $\Delta_{d}(i)$ are matching those of the GIMT calculations. This allows us to compare the number of zero-energy states in the uniform phase state, which is the purpose of the SM work. However, the phase crystal state become extremely weak in this alt-BdG setup, with also a large $s$-wave component, making the edge region being almost instead in a $d+is$-wave state. This result in fact illustrates a common feature of our BdG calculations, that they are surprisingly hard to converge into a phase crystal state. Including strong correlations, this problem completely disappears and the phase crystal is stable in a large parameter regime. 

{\noindent {\textbf {Acknowledgments.}}} We thank P.~Holmvall for useful discussions. We gratefully acknowledge financial support from the Swedish Research Council (Vetenskapsr\aa det, Grant No.~2018-03488) and the Knut and Alice Wallenberg Foundation through the Wallenberg Academy Fellows program. The computations were enabled by resources provided by the Swedish National Infrastructure for Computing (SNIC) at the Uppsala Multidisciplinary Center for Advanced Computational Science (UPPMAX) partially funded by the Swedish Research Council through grant agreement no.~2018-05973.

{\noindent {\textbf {Author contributions.}}} A.B.S.~and D.C.~conceived the project. D.C.~performed the numerical simulations. All authors analyzed and interpreted the results. D.C.~and A.B.S.~wrote the manuscript with inputs from all authors. 

{\noindent {\textbf {Competing interests.}}} The authors declare no competing financial or non-financial interests. 

{\noindent {\textbf {Data availability.}}} The data are available from the corresponding author upon reasonable request. 

{\noindent {\textbf {Code availability.}}} The codes are available from the corresponding author upon reasonable request. 

 \bibliographystyle{apsrev4-1}
\bibliography{Cuprates}

\begin{thebibliography}{69}%
\makeatletter
\providecommand \@ifxundefined [1]{%
 \@ifx{#1\undefined}
}%
\providecommand \@ifnum [1]{%
 \ifnum #1\expandafter \@firstoftwo
 \else \expandafter \@secondoftwo
 \fi
}%
\providecommand \@ifx [1]{%
 \ifx #1\expandafter \@firstoftwo
 \else \expandafter \@secondoftwo
 \fi
}%
\providecommand \natexlab [1]{#1}%
\providecommand \enquote  [1]{``#1''}%
\providecommand \bibnamefont  [1]{#1}%
\providecommand \bibfnamefont [1]{#1}%
\providecommand \citenamefont [1]{#1}%
\providecommand \href@noop [0]{\@secondoftwo}%
\providecommand \href [0]{\begingroup \@sanitize@url \@href}%
\providecommand \@href[1]{\@@startlink{#1}\@@href}%
\providecommand \@@href[1]{\endgroup#1\@@endlink}%
\providecommand \@sanitize@url [0]{\catcode `\\12\catcode `\$12\catcode
  `\&12\catcode `\#12\catcode `\^12\catcode `\_12\catcode `\%12\relax}%
\providecommand \@@startlink[1]{}%
\providecommand \@@endlink[0]{}%
\providecommand \url  [0]{\begingroup\@sanitize@url \@url }%
\providecommand \@url [1]{\endgroup\@href {#1}{\urlprefix }}%
\providecommand \urlprefix  [0]{URL }%
\providecommand \Eprint [0]{\href }%
\providecommand \doibase [0]{http://dx.doi.org/}%
\providecommand \selectlanguage [0]{\@gobble}%
\providecommand \bibinfo  [0]{\@secondoftwo}%
\providecommand \bibfield  [0]{\@secondoftwo}%
\providecommand \translation [1]{[#1]}%
\providecommand \BibitemOpen [0]{}%
\providecommand \bibitemStop [0]{}%
\providecommand \bibitemNoStop [0]{.\EOS\space}%
\providecommand \EOS [0]{\spacefactor3000\relax}%
\providecommand \BibitemShut  [1]{\csname bibitem#1\endcsname}%
\let\auto@bib@innerbib\@empty
\bibitem [{\citenamefont {Lee}\ \emph {et~al.}(2006)\citenamefont {Lee},
  \citenamefont {Nagaosa},\ and\ \citenamefont {Wen}}]{Lee06}%
  \BibitemOpen
  \bibfield  {author} {\bibinfo {author} {\bibfnamefont {P.~A.}\ \bibnamefont
  {Lee}}, \bibinfo {author} {\bibfnamefont {N.}~\bibnamefont {Nagaosa}}, \ and\
  \bibinfo {author} {\bibfnamefont {X.-G.}\ \bibnamefont {Wen}},\ }\href
  {\doibase 10.1103/RevModPhys.78.17} {\bibfield  {journal} {\bibinfo
  {journal} {Rev. Mod. Phys.}\ }\textbf {\bibinfo {volume} {78}},\ \bibinfo
  {pages} {17} (\bibinfo {year} {2006})}\BibitemShut {NoStop}%
\bibitem [{\citenamefont {Paschen}\ and\ \citenamefont {Si}(2021)}]{Paschen21}%
  \BibitemOpen
  \bibfield  {author} {\bibinfo {author} {\bibfnamefont {S.}~\bibnamefont
  {Paschen}}\ and\ \bibinfo {author} {\bibfnamefont {Q.}~\bibnamefont {Si}},\
  }\href {\doibase 10.1038/s42254-020-00262-6} {\bibfield  {journal} {\bibinfo
  {journal} {Nat. Rev. Phys.}\ }\textbf {\bibinfo {volume} {3}},\ \bibinfo
  {pages} {9} (\bibinfo {year} {2021})}\BibitemShut {NoStop}%
\bibitem [{\citenamefont {Anderson}(1958)}]{Anderson58}%
  \BibitemOpen
  \bibfield  {author} {\bibinfo {author} {\bibfnamefont {P.~W.}\ \bibnamefont
  {Anderson}},\ }\href {\doibase 10.1103/PhysRev.109.1492} {\bibfield
  {journal} {\bibinfo  {journal} {Phys. Rev.}\ }\textbf {\bibinfo {volume}
  {109}},\ \bibinfo {pages} {1492} (\bibinfo {year} {1958})}\BibitemShut
  {NoStop}%
\bibitem [{\citenamefont {Goldman}\ and\ \citenamefont
  {Markovic}(1998)}]{Goldman98}%
  \BibitemOpen
  \bibfield  {author} {\bibinfo {author} {\bibfnamefont {A.~M.}\ \bibnamefont
  {Goldman}}\ and\ \bibinfo {author} {\bibfnamefont {N.}~\bibnamefont
  {Markovic}},\ }\href
  {https://physicstoday.scitation.org/doi/10.1063/1.882069} {\bibfield
  {journal} {\bibinfo  {journal} {Phys. Today}\ }\textbf {\bibinfo {volume}
  {51}},\ \bibinfo {pages} {39} (\bibinfo {year} {1998})}\BibitemShut {NoStop}%
\bibitem [{\citenamefont {Moore}(2010)}]{Moore10}%
  \BibitemOpen
  \bibfield  {author} {\bibinfo {author} {\bibfnamefont {J.~E.}\ \bibnamefont
  {Moore}},\ }\href {\doibase 10.1038/nature08916} {\bibfield  {journal}
  {\bibinfo  {journal} {Nature}\ }\textbf {\bibinfo {volume} {464}},\ \bibinfo
  {pages} {194} (\bibinfo {year} {2010})}\BibitemShut {NoStop}%
\bibitem [{\citenamefont {Sato}\ and\ \citenamefont {Ando}(2017)}]{Sato17}%
  \BibitemOpen
  \bibfield  {author} {\bibinfo {author} {\bibfnamefont {M.}~\bibnamefont
  {Sato}}\ and\ \bibinfo {author} {\bibfnamefont {Y.}~\bibnamefont {Ando}},\
  }\href {\doibase 10.1088/1361-6633/aa6ac7} {\bibfield  {journal} {\bibinfo
  {journal} {Rep. Prog. Phys.}\ }\textbf {\bibinfo {volume} {80}},\ \bibinfo
  {pages} {076501} (\bibinfo {year} {2017})}\BibitemShut {NoStop}%
\bibitem [{\citenamefont {Punnoose}\ and\ \citenamefont
  {Finkel{\textquoteright}stein}(2005)}]{Punnoose05}%
  \BibitemOpen
  \bibfield  {author} {\bibinfo {author} {\bibfnamefont {A.}~\bibnamefont
  {Punnoose}}\ and\ \bibinfo {author} {\bibfnamefont {A.~M.}\ \bibnamefont
  {Finkel{\textquoteright}stein}},\ }\href {\doibase 10.1126/science.1115660}
  {\bibfield  {journal} {\bibinfo  {journal} {Science}\ }\textbf {\bibinfo
  {volume} {310}},\ \bibinfo {pages} {289} (\bibinfo {year}
  {2005})}\BibitemShut {NoStop}%
\bibitem [{\citenamefont {Meier}\ \emph {et~al.}(2018)\citenamefont {Meier},
  \citenamefont {An}, \citenamefont {Dauphin}, \citenamefont {Maffei},
  \citenamefont {Massignan}, \citenamefont {Hughes},\ and\ \citenamefont
  {Gadway}}]{Meier18}%
  \BibitemOpen
  \bibfield  {author} {\bibinfo {author} {\bibfnamefont {E.~J.}\ \bibnamefont
  {Meier}}, \bibinfo {author} {\bibfnamefont {F.~A.}\ \bibnamefont {An}},
  \bibinfo {author} {\bibfnamefont {A.}~\bibnamefont {Dauphin}}, \bibinfo
  {author} {\bibfnamefont {M.}~\bibnamefont {Maffei}}, \bibinfo {author}
  {\bibfnamefont {P.}~\bibnamefont {Massignan}}, \bibinfo {author}
  {\bibfnamefont {T.~L.}\ \bibnamefont {Hughes}}, \ and\ \bibinfo {author}
  {\bibfnamefont {B.}~\bibnamefont {Gadway}},\ }\href {\doibase
  10.1126/science.aat3406} {\bibfield  {journal} {\bibinfo  {journal}
  {Science}\ }\textbf {\bibinfo {volume} {362}},\ \bibinfo {pages} {929}
  (\bibinfo {year} {2018})}\BibitemShut {NoStop}%
\bibitem [{\citenamefont {Shi}\ \emph {et~al.}(2021)\citenamefont {Shi},
  \citenamefont {Wieder}, \citenamefont {Meyerheim}, \citenamefont {Sun},
  \citenamefont {Zhang}, \citenamefont {Li}, \citenamefont {Shen},
  \citenamefont {Qi}, \citenamefont {Yang}, \citenamefont {Jena}, \citenamefont
  {Werner}, \citenamefont {Koepernik}, \citenamefont {Parkin}, \citenamefont
  {Chen}, \citenamefont {Felser}, \citenamefont {Bernevig},\ and\ \citenamefont
  {Wang}}]{Shi21}%
  \BibitemOpen
  \bibfield  {author} {\bibinfo {author} {\bibfnamefont {W.}~\bibnamefont
  {Shi}}, \bibinfo {author} {\bibfnamefont {B.~J.}\ \bibnamefont {Wieder}},
  \bibinfo {author} {\bibfnamefont {H.~L.}\ \bibnamefont {Meyerheim}}, \bibinfo
  {author} {\bibfnamefont {Y.}~\bibnamefont {Sun}}, \bibinfo {author}
  {\bibfnamefont {Y.}~\bibnamefont {Zhang}}, \bibinfo {author} {\bibfnamefont
  {Y.}~\bibnamefont {Li}}, \bibinfo {author} {\bibfnamefont {L.}~\bibnamefont
  {Shen}}, \bibinfo {author} {\bibfnamefont {Y.}~\bibnamefont {Qi}}, \bibinfo
  {author} {\bibfnamefont {L.}~\bibnamefont {Yang}}, \bibinfo {author}
  {\bibfnamefont {J.}~\bibnamefont {Jena}}, \bibinfo {author} {\bibfnamefont
  {P.}~\bibnamefont {Werner}}, \bibinfo {author} {\bibfnamefont
  {K.}~\bibnamefont {Koepernik}}, \bibinfo {author} {\bibfnamefont
  {S.}~\bibnamefont {Parkin}}, \bibinfo {author} {\bibfnamefont
  {Y.}~\bibnamefont {Chen}}, \bibinfo {author} {\bibfnamefont {C.}~\bibnamefont
  {Felser}}, \bibinfo {author} {\bibfnamefont {B.~A.}\ \bibnamefont
  {Bernevig}}, \ and\ \bibinfo {author} {\bibfnamefont {Z.}~\bibnamefont
  {Wang}},\ }\href {\doibase 10.1038/s41567-020-01104-z} {\bibfield  {journal}
  {\bibinfo  {journal} {Nat. Phys.}\ }\textbf {\bibinfo {volume} {17}},\
  \bibinfo {pages} {381} (\bibinfo {year} {2021})}\BibitemShut {NoStop}%
\bibitem [{\citenamefont {Kashiwaya}\ and\ \citenamefont
  {Tanaka}(2000)}]{Kashiwaya00}%
  \BibitemOpen
  \bibfield  {author} {\bibinfo {author} {\bibfnamefont {S.}~\bibnamefont
  {Kashiwaya}}\ and\ \bibinfo {author} {\bibfnamefont {Y.}~\bibnamefont
  {Tanaka}},\ }\href {\doibase 10.1088/0034-4885/63/10/202} {\bibfield
  {journal} {\bibinfo  {journal} {Rep. Prog. Phys.}\ }\textbf {\bibinfo
  {volume} {63}},\ \bibinfo {pages} {1641} (\bibinfo {year}
  {2000})}\BibitemShut {NoStop}%
\bibitem [{\citenamefont {L\"{o}fwander}\ \emph {et~al.}(2001)\citenamefont
  {L\"{o}fwander}, \citenamefont {Shumeiko},\ and\ \citenamefont
  {Wendin}}]{Lofwander01}%
  \BibitemOpen
  \bibfield  {author} {\bibinfo {author} {\bibfnamefont {T.}~\bibnamefont
  {L\"{o}fwander}}, \bibinfo {author} {\bibfnamefont {V.~S.}\ \bibnamefont
  {Shumeiko}}, \ and\ \bibinfo {author} {\bibfnamefont {G.}~\bibnamefont
  {Wendin}},\ }\href {\doibase 10.1088/0953-2048/14/5/201} {\bibfield
  {journal} {\bibinfo  {journal} {Supercond. Sci. Technol.}\ }\textbf {\bibinfo
  {volume} {14}},\ \bibinfo {pages} {R53} (\bibinfo {year} {2001})}\BibitemShut
  {NoStop}%
\bibitem [{\citenamefont {Ryu}\ and\ \citenamefont {Hatsugai}(2002)}]{Ryu02}%
  \BibitemOpen
  \bibfield  {author} {\bibinfo {author} {\bibfnamefont {S.}~\bibnamefont
  {Ryu}}\ and\ \bibinfo {author} {\bibfnamefont {Y.}~\bibnamefont {Hatsugai}},\
  }\href {\doibase 10.1103/PhysRevLett.89.077002} {\bibfield  {journal}
  {\bibinfo  {journal} {Phys. Rev. Lett.}\ }\textbf {\bibinfo {volume} {89}},\
  \bibinfo {pages} {077002} (\bibinfo {year} {2002})}\BibitemShut {NoStop}%
\bibitem [{\citenamefont {Sato}\ \emph {et~al.}(2011)\citenamefont {Sato},
  \citenamefont {Tanaka}, \citenamefont {Yada},\ and\ \citenamefont
  {Yokoyama}}]{Sato11}%
  \BibitemOpen
  \bibfield  {author} {\bibinfo {author} {\bibfnamefont {M.}~\bibnamefont
  {Sato}}, \bibinfo {author} {\bibfnamefont {Y.}~\bibnamefont {Tanaka}},
  \bibinfo {author} {\bibfnamefont {K.}~\bibnamefont {Yada}}, \ and\ \bibinfo
  {author} {\bibfnamefont {T.}~\bibnamefont {Yokoyama}},\ }\href {\doibase
  10.1103/PhysRevB.83.224511} {\bibfield  {journal} {\bibinfo  {journal} {Phys.
  Rev. B}\ }\textbf {\bibinfo {volume} {83}},\ \bibinfo {pages} {224511}
  (\bibinfo {year} {2011})}\BibitemShut {NoStop}%
\bibitem [{\citenamefont {Howald}\ \emph {et~al.}(2001)\citenamefont {Howald},
  \citenamefont {Fournier},\ and\ \citenamefont {Kapitulnik}}]{Howald01}%
  \BibitemOpen
  \bibfield  {author} {\bibinfo {author} {\bibfnamefont {C.}~\bibnamefont
  {Howald}}, \bibinfo {author} {\bibfnamefont {P.}~\bibnamefont {Fournier}}, \
  and\ \bibinfo {author} {\bibfnamefont {A.}~\bibnamefont {Kapitulnik}},\
  }\href {\doibase 10.1103/PhysRevB.64.100504} {\bibfield  {journal} {\bibinfo
  {journal} {Phys. Rev. B}\ }\textbf {\bibinfo {volume} {64}},\ \bibinfo
  {pages} {100504} (\bibinfo {year} {2001})}\BibitemShut {NoStop}%
\bibitem [{\citenamefont {Alloul}\ \emph {et~al.}(2009)\citenamefont {Alloul},
  \citenamefont {Bobroff}, \citenamefont {Gabay},\ and\ \citenamefont
  {Hirschfeld}}]{Alloul09}%
  \BibitemOpen
  \bibfield  {author} {\bibinfo {author} {\bibfnamefont {H.}~\bibnamefont
  {Alloul}}, \bibinfo {author} {\bibfnamefont {J.}~\bibnamefont {Bobroff}},
  \bibinfo {author} {\bibfnamefont {M.}~\bibnamefont {Gabay}}, \ and\ \bibinfo
  {author} {\bibfnamefont {P.~J.}\ \bibnamefont {Hirschfeld}},\ }\href
  {\doibase 10.1103/RevModPhys.81.45} {\bibfield  {journal} {\bibinfo
  {journal} {Rev. Mod. Phys.}\ }\textbf {\bibinfo {volume} {81}},\ \bibinfo
  {pages} {45} (\bibinfo {year} {2009})}\BibitemShut {NoStop}%
\bibitem [{\citenamefont {Potter}\ and\ \citenamefont {Lee}(2014)}]{Potter14}%
  \BibitemOpen
  \bibfield  {author} {\bibinfo {author} {\bibfnamefont {A.~C.}\ \bibnamefont
  {Potter}}\ and\ \bibinfo {author} {\bibfnamefont {P.~A.}\ \bibnamefont
  {Lee}},\ }\href {\doibase 10.1103/PhysRevLett.112.117002} {\bibfield
  {journal} {\bibinfo  {journal} {Phys. Rev. Lett.}\ }\textbf {\bibinfo
  {volume} {112}},\ \bibinfo {pages} {117002} (\bibinfo {year}
  {2014})}\BibitemShut {NoStop}%
\bibitem [{\citenamefont {{Honerkamp, C.}}\ \emph {et~al.}(2000)\citenamefont
  {{Honerkamp, C.}}, \citenamefont {{Wakabayashi, K.}},\ and\ \citenamefont
  {{Sigrist, M.}}}]{Honerkamp00}%
  \BibitemOpen
  \bibfield  {author} {\bibinfo {author} {\bibnamefont {{Honerkamp, C.}}},
  \bibinfo {author} {\bibnamefont {{Wakabayashi, K.}}}, \ and\ \bibinfo
  {author} {\bibnamefont {{Sigrist, M.}}},\ }\href {\doibase
  10.1209/epl/i2000-00280-2} {\bibfield  {journal} {\bibinfo  {journal}
  {Europhys. Lett.}\ }\textbf {\bibinfo {volume} {50}},\ \bibinfo {pages} {368}
  (\bibinfo {year} {2000})}\BibitemShut {NoStop}%
\bibitem [{\citenamefont {L\"ofwander}\ \emph {et~al.}(2000)\citenamefont
  {L\"ofwander}, \citenamefont {Shumeiko},\ and\ \citenamefont
  {Wendin}}]{Lofwander00}%
  \BibitemOpen
  \bibfield  {author} {\bibinfo {author} {\bibfnamefont {T.}~\bibnamefont
  {L\"ofwander}}, \bibinfo {author} {\bibfnamefont {V.~S.}\ \bibnamefont
  {Shumeiko}}, \ and\ \bibinfo {author} {\bibfnamefont {G.}~\bibnamefont
  {Wendin}},\ }\href {\doibase 10.1103/PhysRevB.62.R14653} {\bibfield
  {journal} {\bibinfo  {journal} {Phys. Rev. B}\ }\textbf {\bibinfo {volume}
  {62}},\ \bibinfo {pages} {R14653} (\bibinfo {year} {2000})}\BibitemShut
  {NoStop}%
\bibitem [{\citenamefont {Black-Schaffer}\ \emph {et~al.}(2013)\citenamefont
  {Black-Schaffer}, \citenamefont {Golubev}, \citenamefont {Bauch},
  \citenamefont {Lombardi},\ and\ \citenamefont
  {Fogelstr\"om}}]{Black-Schaffer13}%
  \BibitemOpen
  \bibfield  {author} {\bibinfo {author} {\bibfnamefont {A.~M.}\ \bibnamefont
  {Black-Schaffer}}, \bibinfo {author} {\bibfnamefont {D.~S.}\ \bibnamefont
  {Golubev}}, \bibinfo {author} {\bibfnamefont {T.}~\bibnamefont {Bauch}},
  \bibinfo {author} {\bibfnamefont {F.}~\bibnamefont {Lombardi}}, \ and\
  \bibinfo {author} {\bibfnamefont {M.}~\bibnamefont {Fogelstr\"om}},\ }\href
  {\doibase 10.1103/PhysRevLett.110.197001} {\bibfield  {journal} {\bibinfo
  {journal} {Phys. Rev. Lett.}\ }\textbf {\bibinfo {volume} {110}},\ \bibinfo
  {pages} {197001} (\bibinfo {year} {2013})}\BibitemShut {NoStop}%
\bibitem [{\citenamefont {Nagai}\ \emph {et~al.}(2017)\citenamefont {Nagai},
  \citenamefont {Ota},\ and\ \citenamefont {Tanaka}}]{Nagai17}%
  \BibitemOpen
  \bibfield  {author} {\bibinfo {author} {\bibfnamefont {Y.}~\bibnamefont
  {Nagai}}, \bibinfo {author} {\bibfnamefont {Y.}~\bibnamefont {Ota}}, \ and\
  \bibinfo {author} {\bibfnamefont {K.}~\bibnamefont {Tanaka}},\ }\href
  {\doibase 10.1103/PhysRevB.96.060503} {\bibfield  {journal} {\bibinfo
  {journal} {Phys. Rev. B}\ }\textbf {\bibinfo {volume} {96}},\ \bibinfo
  {pages} {060503} (\bibinfo {year} {2017})}\BibitemShut {NoStop}%
\bibitem [{\citenamefont {Matsubara}\ and\ \citenamefont
  {Kontani}(2020)}]{Matsubara20}%
  \BibitemOpen
  \bibfield  {author} {\bibinfo {author} {\bibfnamefont {S.}~\bibnamefont
  {Matsubara}}\ and\ \bibinfo {author} {\bibfnamefont {H.}~\bibnamefont
  {Kontani}},\ }\href {\doibase 10.1103/PhysRevB.101.235103} {\bibfield
  {journal} {\bibinfo  {journal} {Phys. Rev. B}\ }\textbf {\bibinfo {volume}
  {101}},\ \bibinfo {pages} {235103} (\bibinfo {year} {2020})}\BibitemShut
  {NoStop}%
\bibitem [{\citenamefont {H{\aa}kansson}\ \emph {et~al.}(2015)\citenamefont
  {H{\aa}kansson}, \citenamefont {L{\"o}fwander},\ and\ \citenamefont
  {Fogelstr{\"o}m}}]{Hakansson2015}%
  \BibitemOpen
  \bibfield  {author} {\bibinfo {author} {\bibfnamefont {M.}~\bibnamefont
  {H{\aa}kansson}}, \bibinfo {author} {\bibfnamefont {T.}~\bibnamefont
  {L{\"o}fwander}}, \ and\ \bibinfo {author} {\bibfnamefont {M.}~\bibnamefont
  {Fogelstr{\"o}m}},\ }\href {\doibase 10.1038/nphys3383} {\bibfield  {journal}
  {\bibinfo  {journal} {Nat. Phys.}\ }\textbf {\bibinfo {volume} {11}},\
  \bibinfo {pages} {755} (\bibinfo {year} {2015})}\BibitemShut {NoStop}%
\bibitem [{\citenamefont {Holmvall}\ \emph {et~al.}(2018)\citenamefont
  {Holmvall}, \citenamefont {Vorontsov}, \citenamefont {Fogelstr{\"o}m},\ and\
  \citenamefont {L{\"o}fwander}}]{Holmvall18}%
  \BibitemOpen
  \bibfield  {author} {\bibinfo {author} {\bibfnamefont {P.}~\bibnamefont
  {Holmvall}}, \bibinfo {author} {\bibfnamefont {A.~B.}\ \bibnamefont
  {Vorontsov}}, \bibinfo {author} {\bibfnamefont {M.}~\bibnamefont
  {Fogelstr{\"o}m}}, \ and\ \bibinfo {author} {\bibfnamefont {T.}~\bibnamefont
  {L{\"o}fwander}},\ }\href {\doibase 10.1038/s41467-018-04531-y} {\bibfield
  {journal} {\bibinfo  {journal} {Nat. Commun.}\ }\textbf {\bibinfo {volume}
  {9}},\ \bibinfo {pages} {2190} (\bibinfo {year} {2018})}\BibitemShut
  {NoStop}%
\bibitem [{\citenamefont {Holmvall}\ \emph {et~al.}(2020)\citenamefont
  {Holmvall}, \citenamefont {Fogelstr\"om}, \citenamefont {L\"ofwander},\ and\
  \citenamefont {Vorontsov}}]{Holmvall20}%
  \BibitemOpen
  \bibfield  {author} {\bibinfo {author} {\bibfnamefont {P.}~\bibnamefont
  {Holmvall}}, \bibinfo {author} {\bibfnamefont {M.}~\bibnamefont
  {Fogelstr\"om}}, \bibinfo {author} {\bibfnamefont {T.}~\bibnamefont
  {L\"ofwander}}, \ and\ \bibinfo {author} {\bibfnamefont {A.~B.}\ \bibnamefont
  {Vorontsov}},\ }\href {\doibase 10.1103/PhysRevResearch.2.013104} {\bibfield
  {journal} {\bibinfo  {journal} {Phys. Rev. Research}\ }\textbf {\bibinfo
  {volume} {2}},\ \bibinfo {pages} {013104} (\bibinfo {year}
  {2020})}\BibitemShut {NoStop}%
\bibitem [{\citenamefont {Wennerdal}\ \emph {et~al.}(2020)\citenamefont
  {Wennerdal}, \citenamefont {Ask}, \citenamefont {Holmvall}, \citenamefont
  {L\"ofwander},\ and\ \citenamefont {Fogelstr\"om}}]{Wennerdal20}%
  \BibitemOpen
  \bibfield  {author} {\bibinfo {author} {\bibfnamefont {N.~W.}\ \bibnamefont
  {Wennerdal}}, \bibinfo {author} {\bibfnamefont {A.}~\bibnamefont {Ask}},
  \bibinfo {author} {\bibfnamefont {P.}~\bibnamefont {Holmvall}}, \bibinfo
  {author} {\bibfnamefont {T.}~\bibnamefont {L\"ofwander}}, \ and\ \bibinfo
  {author} {\bibfnamefont {M.}~\bibnamefont {Fogelstr\"om}},\ }\href {\doibase
  10.1103/PhysRevResearch.2.043198} {\bibfield  {journal} {\bibinfo  {journal}
  {Phys. Rev. Research}\ }\textbf {\bibinfo {volume} {2}},\ \bibinfo {pages}
  {043198} (\bibinfo {year} {2020})}\BibitemShut {NoStop}%
\bibitem [{\citenamefont {Kalenkov}\ \emph {et~al.}(2004)\citenamefont
  {Kalenkov}, \citenamefont {Fogelstr\"om},\ and\ \citenamefont
  {Barash}}]{Kalenkov04}%
  \BibitemOpen
  \bibfield  {author} {\bibinfo {author} {\bibfnamefont {M.~S.}\ \bibnamefont
  {Kalenkov}}, \bibinfo {author} {\bibfnamefont {M.}~\bibnamefont
  {Fogelstr\"om}}, \ and\ \bibinfo {author} {\bibfnamefont {Y.~S.}\
  \bibnamefont {Barash}},\ }\href {\doibase 10.1103/PhysRevB.70.184505}
  {\bibfield  {journal} {\bibinfo  {journal} {Phys. Rev. B}\ }\textbf {\bibinfo
  {volume} {70}},\ \bibinfo {pages} {184505} (\bibinfo {year}
  {2004})}\BibitemShut {NoStop}%
\bibitem [{\citenamefont {Ikegaya}\ and\ \citenamefont
  {Asano}(2017)}]{Ikegaya17}%
  \BibitemOpen
  \bibfield  {author} {\bibinfo {author} {\bibfnamefont {S.}~\bibnamefont
  {Ikegaya}}\ and\ \bibinfo {author} {\bibfnamefont {Y.}~\bibnamefont
  {Asano}},\ }\href {\doibase 10.1103/PhysRevB.95.214503} {\bibfield  {journal}
  {\bibinfo  {journal} {Phys. Rev. B}\ }\textbf {\bibinfo {volume} {95}},\
  \bibinfo {pages} {214503} (\bibinfo {year} {2017})}\BibitemShut {NoStop}%
\bibitem [{\citenamefont {Covington}\ \emph {et~al.}(1997)\citenamefont
  {Covington}, \citenamefont {Aprili}, \citenamefont {Paraoanu}, \citenamefont
  {Greene}, \citenamefont {Xu}, \citenamefont {Zhu},\ and\ \citenamefont
  {Mirkin}}]{Covington97}%
  \BibitemOpen
  \bibfield  {author} {\bibinfo {author} {\bibfnamefont {M.}~\bibnamefont
  {Covington}}, \bibinfo {author} {\bibfnamefont {M.}~\bibnamefont {Aprili}},
  \bibinfo {author} {\bibfnamefont {E.}~\bibnamefont {Paraoanu}}, \bibinfo
  {author} {\bibfnamefont {L.~H.}\ \bibnamefont {Greene}}, \bibinfo {author}
  {\bibfnamefont {F.}~\bibnamefont {Xu}}, \bibinfo {author} {\bibfnamefont
  {J.}~\bibnamefont {Zhu}}, \ and\ \bibinfo {author} {\bibfnamefont {C.~A.}\
  \bibnamefont {Mirkin}},\ }\href {\doibase 10.1103/PhysRevLett.79.277}
  {\bibfield  {journal} {\bibinfo  {journal} {Phys. Rev. Lett.}\ }\textbf
  {\bibinfo {volume} {79}},\ \bibinfo {pages} {277} (\bibinfo {year}
  {1997})}\BibitemShut {NoStop}%
\bibitem [{\citenamefont {Dagan}\ and\ \citenamefont
  {Deutscher}(2001)}]{Dagan01}%
  \BibitemOpen
  \bibfield  {author} {\bibinfo {author} {\bibfnamefont {Y.}~\bibnamefont
  {Dagan}}\ and\ \bibinfo {author} {\bibfnamefont {G.}~\bibnamefont
  {Deutscher}},\ }\href {\doibase 10.1103/PhysRevLett.87.177004} {\bibfield
  {journal} {\bibinfo  {journal} {Phys. Rev. Lett.}\ }\textbf {\bibinfo
  {volume} {87}},\ \bibinfo {pages} {177004} (\bibinfo {year}
  {2001})}\BibitemShut {NoStop}%
\bibitem [{\citenamefont {Gustafsson}\ \emph {et~al.}(2013)\citenamefont
  {Gustafsson}, \citenamefont {Golubev}, \citenamefont {Fogelstr{\"o}m},
  \citenamefont {Claeson}, \citenamefont {Kubatkin}, \citenamefont {Bauch},\
  and\ \citenamefont {Lombardi}}]{Gustafsson13}%
  \BibitemOpen
  \bibfield  {author} {\bibinfo {author} {\bibfnamefont {D.}~\bibnamefont
  {Gustafsson}}, \bibinfo {author} {\bibfnamefont {D.}~\bibnamefont {Golubev}},
  \bibinfo {author} {\bibfnamefont {M.}~\bibnamefont {Fogelstr{\"o}m}},
  \bibinfo {author} {\bibfnamefont {T.}~\bibnamefont {Claeson}}, \bibinfo
  {author} {\bibfnamefont {S.}~\bibnamefont {Kubatkin}}, \bibinfo {author}
  {\bibfnamefont {T.}~\bibnamefont {Bauch}}, \ and\ \bibinfo {author}
  {\bibfnamefont {F.}~\bibnamefont {Lombardi}},\ }\href {\doibase
  10.1038/nnano.2012.214} {\bibfield  {journal} {\bibinfo  {journal} {Nat.
  Nanotechnol.}\ }\textbf {\bibinfo {volume} {8}},\ \bibinfo {pages} {25}
  (\bibinfo {year} {2013})}\BibitemShut {NoStop}%
\bibitem [{\citenamefont {Alff}\ \emph {et~al.}(1997)\citenamefont {Alff},
  \citenamefont {Takashima}, \citenamefont {Kashiwaya}, \citenamefont {Terada},
  \citenamefont {Ihara}, \citenamefont {Tanaka}, \citenamefont {Koyanagi},\
  and\ \citenamefont {Kajimura}}]{Alff97}%
  \BibitemOpen
  \bibfield  {author} {\bibinfo {author} {\bibfnamefont {L.}~\bibnamefont
  {Alff}}, \bibinfo {author} {\bibfnamefont {H.}~\bibnamefont {Takashima}},
  \bibinfo {author} {\bibfnamefont {S.}~\bibnamefont {Kashiwaya}}, \bibinfo
  {author} {\bibfnamefont {N.}~\bibnamefont {Terada}}, \bibinfo {author}
  {\bibfnamefont {H.}~\bibnamefont {Ihara}}, \bibinfo {author} {\bibfnamefont
  {Y.}~\bibnamefont {Tanaka}}, \bibinfo {author} {\bibfnamefont
  {M.}~\bibnamefont {Koyanagi}}, \ and\ \bibinfo {author} {\bibfnamefont
  {K.}~\bibnamefont {Kajimura}},\ }\href {\doibase 10.1103/PhysRevB.55.R14757}
  {\bibfield  {journal} {\bibinfo  {journal} {Phys. Rev. B}\ }\textbf {\bibinfo
  {volume} {55}},\ \bibinfo {pages} {R14757} (\bibinfo {year}
  {1997})}\BibitemShut {NoStop}%
\bibitem [{\citenamefont {Neils}\ and\ \citenamefont
  {Van~Harlingen}(2002)}]{Neils02}%
  \BibitemOpen
  \bibfield  {author} {\bibinfo {author} {\bibfnamefont {W.~K.}\ \bibnamefont
  {Neils}}\ and\ \bibinfo {author} {\bibfnamefont {D.~J.}\ \bibnamefont
  {Van~Harlingen}},\ }\href {\doibase 10.1103/PhysRevLett.88.047001} {\bibfield
   {journal} {\bibinfo  {journal} {Phys. Rev. Lett.}\ }\textbf {\bibinfo
  {volume} {88}},\ \bibinfo {pages} {047001} (\bibinfo {year}
  {2002})}\BibitemShut {NoStop}%
\bibitem [{\citenamefont {Scalapino}(1995)}]{Scalapino95}%
  \BibitemOpen
  \bibfield  {author} {\bibinfo {author} {\bibfnamefont {D.}~\bibnamefont
  {Scalapino}},\ }\href {\doibase https://doi.org/10.1016/0370-1573(94)00086-I}
  {\bibfield  {journal} {\bibinfo  {journal} {Physics Reports}\ }\textbf
  {\bibinfo {volume} {250}},\ \bibinfo {pages} {329 } (\bibinfo {year}
  {1995})}\BibitemShut {NoStop}%
\bibitem [{\citenamefont {Anderson}\ \emph {et~al.}(2004)\citenamefont
  {Anderson}, \citenamefont {Lee}, \citenamefont {Randeria}, \citenamefont
  {Rice}, \citenamefont {Trivedi},\ and\ \citenamefont {Zhang}}]{Anderson04}%
  \BibitemOpen
  \bibfield  {author} {\bibinfo {author} {\bibfnamefont {P.~W.}\ \bibnamefont
  {Anderson}}, \bibinfo {author} {\bibfnamefont {P.~A.}\ \bibnamefont {Lee}},
  \bibinfo {author} {\bibfnamefont {M.}~\bibnamefont {Randeria}}, \bibinfo
  {author} {\bibfnamefont {T.~M.}\ \bibnamefont {Rice}}, \bibinfo {author}
  {\bibfnamefont {N.}~\bibnamefont {Trivedi}}, \ and\ \bibinfo {author}
  {\bibfnamefont {F.~C.}\ \bibnamefont {Zhang}},\ }\href {\doibase
  10.1088/0953-8984/16/24/R02} {\bibfield  {journal} {\bibinfo  {journal} {J.
  Phys. Condens. Matter}\ }\textbf {\bibinfo {volume} {16}},\ \bibinfo {pages}
  {R755} (\bibinfo {year} {2004})}\BibitemShut {NoStop}%
\bibitem [{\citenamefont {Zhang}\ \emph {et~al.}(1988)\citenamefont {Zhang},
  \citenamefont {Gros}, \citenamefont {Rice},\ and\ \citenamefont
  {Shiba}}]{Zhang88}%
  \BibitemOpen
  \bibfield  {author} {\bibinfo {author} {\bibfnamefont {F.~C.}\ \bibnamefont
  {Zhang}}, \bibinfo {author} {\bibfnamefont {C.}~\bibnamefont {Gros}},
  \bibinfo {author} {\bibfnamefont {T.~M.}\ \bibnamefont {Rice}}, \ and\
  \bibinfo {author} {\bibfnamefont {H.}~\bibnamefont {Shiba}},\ }\href
  {\doibase 10.1088/0953-2048/1/1/009} {\bibfield  {journal} {\bibinfo
  {journal} {Supercond. Sci. Technol.}\ }\textbf {\bibinfo {volume} {1}},\
  \bibinfo {pages} {36} (\bibinfo {year} {1988})}\BibitemShut {NoStop}%
\bibitem [{\citenamefont {Paramekanti}\ \emph {et~al.}(2001)\citenamefont
  {Paramekanti}, \citenamefont {Randeria},\ and\ \citenamefont
  {Trivedi}}]{Paramekanti01}%
  \BibitemOpen
  \bibfield  {author} {\bibinfo {author} {\bibfnamefont {A.}~\bibnamefont
  {Paramekanti}}, \bibinfo {author} {\bibfnamefont {M.}~\bibnamefont
  {Randeria}}, \ and\ \bibinfo {author} {\bibfnamefont {N.}~\bibnamefont
  {Trivedi}},\ }\href {\doibase 10.1103/PhysRevLett.87.217002} {\bibfield
  {journal} {\bibinfo  {journal} {Phys. Rev. Lett.}\ }\textbf {\bibinfo
  {volume} {87}},\ \bibinfo {pages} {217002} (\bibinfo {year}
  {2001})}\BibitemShut {NoStop}%
\bibitem [{\citenamefont {Sensarma}\ \emph {et~al.}(2007)\citenamefont
  {Sensarma}, \citenamefont {Randeria},\ and\ \citenamefont
  {Trivedi}}]{Sensarma07}%
  \BibitemOpen
  \bibfield  {author} {\bibinfo {author} {\bibfnamefont {R.}~\bibnamefont
  {Sensarma}}, \bibinfo {author} {\bibfnamefont {M.}~\bibnamefont {Randeria}},
  \ and\ \bibinfo {author} {\bibfnamefont {N.}~\bibnamefont {Trivedi}},\ }\href
  {\doibase 10.1103/PhysRevLett.98.027004} {\bibfield  {journal} {\bibinfo
  {journal} {Phys. Rev. Lett.}\ }\textbf {\bibinfo {volume} {98}},\ \bibinfo
  {pages} {027004} (\bibinfo {year} {2007})}\BibitemShut {NoStop}%
\bibitem [{\citenamefont {Fukushima}(2008)}]{Fukushima08}%
  \BibitemOpen
  \bibfield  {author} {\bibinfo {author} {\bibfnamefont {N.}~\bibnamefont
  {Fukushima}},\ }\href {\doibase 10.1103/PhysRevB.78.115105} {\bibfield
  {journal} {\bibinfo  {journal} {Phys. Rev. B}\ }\textbf {\bibinfo {volume}
  {78}},\ \bibinfo {pages} {115105} (\bibinfo {year} {2008})}\BibitemShut
  {NoStop}%
\bibitem [{\citenamefont {Fukushima}\ \emph {et~al.}(2009)\citenamefont
  {Fukushima}, \citenamefont {Chou},\ and\ \citenamefont {Lee}}]{Fukushima09}%
  \BibitemOpen
  \bibfield  {author} {\bibinfo {author} {\bibfnamefont {N.}~\bibnamefont
  {Fukushima}}, \bibinfo {author} {\bibfnamefont {C.-P.}\ \bibnamefont {Chou}},
  \ and\ \bibinfo {author} {\bibfnamefont {T.~K.}\ \bibnamefont {Lee}},\ }\href
  {\doibase 10.1103/PhysRevB.79.184510} {\bibfield  {journal} {\bibinfo
  {journal} {Phys. Rev. B}\ }\textbf {\bibinfo {volume} {79}},\ \bibinfo
  {pages} {184510} (\bibinfo {year} {2009})}\BibitemShut {NoStop}%
\bibitem [{\citenamefont {Garg}\ \emph {et~al.}(2008)\citenamefont {Garg},
  \citenamefont {Randeria},\ and\ \citenamefont {Trivedi}}]{Garg08}%
  \BibitemOpen
  \bibfield  {author} {\bibinfo {author} {\bibfnamefont {A.}~\bibnamefont
  {Garg}}, \bibinfo {author} {\bibfnamefont {M.}~\bibnamefont {Randeria}}, \
  and\ \bibinfo {author} {\bibfnamefont {N.}~\bibnamefont {Trivedi}},\ }\href
  {\doibase 10.1038/nphys1026} {\bibfield  {journal} {\bibinfo  {journal} {Nat.
  Phys.}\ }\textbf {\bibinfo {volume} {4}},\ \bibinfo {pages} {762} (\bibinfo
  {year} {2008})}\BibitemShut {NoStop}%
\bibitem [{\citenamefont {Chakraborty}\ and\ \citenamefont
  {Ghosal}(2014)}]{Chakraborty14}%
  \BibitemOpen
  \bibfield  {author} {\bibinfo {author} {\bibfnamefont {D.}~\bibnamefont
  {Chakraborty}}\ and\ \bibinfo {author} {\bibfnamefont {A.}~\bibnamefont
  {Ghosal}},\ }\href {\doibase 10.1088/1367-2630/16/10/103018} {\bibfield
  {journal} {\bibinfo  {journal} {New J. Phys.}\ }\textbf {\bibinfo {volume}
  {16}},\ \bibinfo {pages} {103018} (\bibinfo {year} {2014})}\BibitemShut
  {NoStop}%
\bibitem [{\citenamefont {Zhu}(2016)}]{Zhubook}%
  \BibitemOpen
  \bibfield  {author} {\bibinfo {author} {\bibfnamefont {J.}~\bibnamefont
  {Zhu}},\ }\href {https://books.google.se/books?id=Tep6DAAAQBAJ} {\emph
  {\bibinfo {title} {Bogoliubov-de Gennes Method and Its Applications}}},\
  Lecture Notes in Physics\ (\bibinfo  {publisher} {Springer International
  Publishing},\ \bibinfo {year} {2016})\BibitemShut {NoStop}%
\bibitem [{\citenamefont {Sigrist}\ and\ \citenamefont
  {Ueda}(1991)}]{Sigrist91}%
  \BibitemOpen
  \bibfield  {author} {\bibinfo {author} {\bibfnamefont {M.}~\bibnamefont
  {Sigrist}}\ and\ \bibinfo {author} {\bibfnamefont {K.}~\bibnamefont {Ueda}},\
  }\href {\doibase 10.1103/RevModPhys.63.239} {\bibfield  {journal} {\bibinfo
  {journal} {Rev. Mod. Phys.}\ }\textbf {\bibinfo {volume} {63}},\ \bibinfo
  {pages} {239} (\bibinfo {year} {1991})}\BibitemShut {NoStop}%
\bibitem [{\citenamefont {Vorontsov}(2009)}]{Vorontsov09}%
  \BibitemOpen
  \bibfield  {author} {\bibinfo {author} {\bibfnamefont {A.~B.}\ \bibnamefont
  {Vorontsov}},\ }\href {\doibase 10.1103/PhysRevLett.102.177001} {\bibfield
  {journal} {\bibinfo  {journal} {Phys. Rev. Lett.}\ }\textbf {\bibinfo
  {volume} {102}},\ \bibinfo {pages} {177001} (\bibinfo {year}
  {2009})}\BibitemShut {NoStop}%
\bibitem [{\citenamefont {Tang}\ \emph {et~al.}(2015)\citenamefont {Tang},
  \citenamefont {Miranda},\ and\ \citenamefont {Dobrosavljevic}}]{Tang15}%
  \BibitemOpen
  \bibfield  {author} {\bibinfo {author} {\bibfnamefont {S.}~\bibnamefont
  {Tang}}, \bibinfo {author} {\bibfnamefont {E.}~\bibnamefont {Miranda}}, \
  and\ \bibinfo {author} {\bibfnamefont {V.}~\bibnamefont {Dobrosavljevic}},\
  }\href {\doibase 10.1103/PhysRevB.91.020501} {\bibfield  {journal} {\bibinfo
  {journal} {Phys. Rev. B}\ }\textbf {\bibinfo {volume} {91}},\ \bibinfo
  {pages} {020501} (\bibinfo {year} {2015})}\BibitemShut {NoStop}%
\bibitem [{\citenamefont {Nakosai}\ \emph {et~al.}(2013)\citenamefont
  {Nakosai}, \citenamefont {Tanaka},\ and\ \citenamefont
  {Nagaosa}}]{Nakosai13}%
  \BibitemOpen
  \bibfield  {author} {\bibinfo {author} {\bibfnamefont {S.}~\bibnamefont
  {Nakosai}}, \bibinfo {author} {\bibfnamefont {Y.}~\bibnamefont {Tanaka}}, \
  and\ \bibinfo {author} {\bibfnamefont {N.}~\bibnamefont {Nagaosa}},\ }\href
  {\doibase 10.1103/PhysRevB.88.180503} {\bibfield  {journal} {\bibinfo
  {journal} {Phys. Rev. B}\ }\textbf {\bibinfo {volume} {88}},\ \bibinfo
  {pages} {180503} (\bibinfo {year} {2013})}\BibitemShut {NoStop}%
\bibitem [{\citenamefont {Chen}\ and\ \citenamefont {Schnyder}(2015)}]{Chen15}%
  \BibitemOpen
  \bibfield  {author} {\bibinfo {author} {\bibfnamefont {W.}~\bibnamefont
  {Chen}}\ and\ \bibinfo {author} {\bibfnamefont {A.~P.}\ \bibnamefont
  {Schnyder}},\ }\href {\doibase 10.1103/PhysRevB.92.214502} {\bibfield
  {journal} {\bibinfo  {journal} {Phys. Rev. B}\ }\textbf {\bibinfo {volume}
  {92}},\ \bibinfo {pages} {214502} (\bibinfo {year} {2015})}\BibitemShut
  {NoStop}%
\bibitem [{\citenamefont {Baskaran}(2002)}]{Baskaran2002}%
  \BibitemOpen
  \bibfield  {author} {\bibinfo {author} {\bibfnamefont {G.}~\bibnamefont
  {Baskaran}},\ }\href {\doibase 10.1103/PhysRevB.65.212505} {\bibfield
  {journal} {\bibinfo  {journal} {Phys. Rev. B}\ }\textbf {\bibinfo {volume}
  {65}},\ \bibinfo {pages} {212505} (\bibinfo {year} {2002})}\BibitemShut
  {NoStop}%
\bibitem [{\citenamefont {Schmidt}\ \emph {et~al.}(2018)\citenamefont
  {Schmidt}, \citenamefont {Scherer},\ and\ \citenamefont
  {Black-Schaffer}}]{Schmidt2018}%
  \BibitemOpen
  \bibfield  {author} {\bibinfo {author} {\bibfnamefont {J.}~\bibnamefont
  {Schmidt}}, \bibinfo {author} {\bibfnamefont {D.~D.}\ \bibnamefont
  {Scherer}}, \ and\ \bibinfo {author} {\bibfnamefont {A.~M.}\ \bibnamefont
  {Black-Schaffer}},\ }\href {\doibase 10.1103/PhysRevB.97.014504} {\bibfield
  {journal} {\bibinfo  {journal} {Phys. Rev. B}\ }\textbf {\bibinfo {volume}
  {97}},\ \bibinfo {pages} {014504} (\bibinfo {year} {2018})}\BibitemShut
  {NoStop}%
\bibitem [{\citenamefont {Schnyder}\ and\ \citenamefont
  {Brydon}(2015)}]{Schnyder15}%
  \BibitemOpen
  \bibfield  {author} {\bibinfo {author} {\bibfnamefont {A.~P.}\ \bibnamefont
  {Schnyder}}\ and\ \bibinfo {author} {\bibfnamefont {P.~M.~R.}\ \bibnamefont
  {Brydon}},\ }\href {\doibase 10.1088/0953-8984/27/24/243201} {\bibfield
  {journal} {\bibinfo  {journal} {J. Condens. Matter Phys.}\ }\textbf {\bibinfo
  {volume} {27}},\ \bibinfo {pages} {243201} (\bibinfo {year}
  {2015})}\BibitemShut {NoStop}%
\bibitem [{\citenamefont {Tang}\ \emph {et~al.}(2016)\citenamefont {Tang},
  \citenamefont {Dobrosavljevi\ifmmode~\acute{c}\else \'{c}\fi{}},\ and\
  \citenamefont {Miranda}}]{Tang16}%
  \BibitemOpen
  \bibfield  {author} {\bibinfo {author} {\bibfnamefont {S.}~\bibnamefont
  {Tang}}, \bibinfo {author} {\bibfnamefont {V.}~\bibnamefont
  {Dobrosavljevi\ifmmode~\acute{c}\else \'{c}\fi{}}}, \ and\ \bibinfo {author}
  {\bibfnamefont {E.}~\bibnamefont {Miranda}},\ }\href {\doibase
  10.1103/PhysRevB.93.195109} {\bibfield  {journal} {\bibinfo  {journal} {Phys.
  Rev. B}\ }\textbf {\bibinfo {volume} {93}},\ \bibinfo {pages} {195109}
  (\bibinfo {year} {2016})}\BibitemShut {NoStop}%
\bibitem [{\citenamefont {Wei}\ \emph {et~al.}(1998)\citenamefont {Wei},
  \citenamefont {Yeh}, \citenamefont {Garrigus},\ and\ \citenamefont
  {Strasik}}]{Wei98}%
  \BibitemOpen
  \bibfield  {author} {\bibinfo {author} {\bibfnamefont {J.~Y.~T.}\
  \bibnamefont {Wei}}, \bibinfo {author} {\bibfnamefont {N.-C.}\ \bibnamefont
  {Yeh}}, \bibinfo {author} {\bibfnamefont {D.~F.}\ \bibnamefont {Garrigus}}, \
  and\ \bibinfo {author} {\bibfnamefont {M.}~\bibnamefont {Strasik}},\ }\href
  {\doibase 10.1103/PhysRevLett.81.2542} {\bibfield  {journal} {\bibinfo
  {journal} {Phys. Rev. Lett.}\ }\textbf {\bibinfo {volume} {81}},\ \bibinfo
  {pages} {2542} (\bibinfo {year} {1998})}\BibitemShut {NoStop}%
\bibitem [{\citenamefont {Anderson}(1959)}]{Anderson59}%
  \BibitemOpen
  \bibfield  {author} {\bibinfo {author} {\bibfnamefont {P.}~\bibnamefont
  {Anderson}},\ }\href {\doibase https://doi.org/10.1016/0022-3697(59)90036-8}
  {\bibfield  {journal} {\bibinfo  {journal} {J Phys Chem Solids}\ }\textbf
  {\bibinfo {volume} {11}},\ \bibinfo {pages} {26 } (\bibinfo {year}
  {1959})}\BibitemShut {NoStop}%
\bibitem [{\citenamefont {Ghosal}\ \emph {et~al.}(2018)\citenamefont {Ghosal},
  \citenamefont {Chakraborty},\ and\ \citenamefont {Kaushal}}]{Ghosal18}%
  \BibitemOpen
  \bibfield  {author} {\bibinfo {author} {\bibfnamefont {A.}~\bibnamefont
  {Ghosal}}, \bibinfo {author} {\bibfnamefont {D.}~\bibnamefont {Chakraborty}},
  \ and\ \bibinfo {author} {\bibfnamefont {N.}~\bibnamefont {Kaushal}},\ }\href
  {\doibase https://doi.org/10.1016/j.physb.2017.08.040} {\bibfield  {journal}
  {\bibinfo  {journal} {Physica B Condens. Matter}\ }\textbf {\bibinfo {volume}
  {536}},\ \bibinfo {pages} {867 } (\bibinfo {year} {2018})}\BibitemShut
  {NoStop}%
\bibitem [{\citenamefont {Zhang}\ \emph {et~al.}(2020)\citenamefont {Zhang},
  \citenamefont {Yang},\ and\ \citenamefont {Zhang}}]{Zhang20}%
  \BibitemOpen
  \bibfield  {author} {\bibinfo {author} {\bibfnamefont {G.-M.}\ \bibnamefont
  {Zhang}}, \bibinfo {author} {\bibfnamefont {Y.-f.}\ \bibnamefont {Yang}}, \
  and\ \bibinfo {author} {\bibfnamefont {F.-C.}\ \bibnamefont {Zhang}},\ }\href
  {\doibase 10.1103/PhysRevB.101.020501} {\bibfield  {journal} {\bibinfo
  {journal} {Phys. Rev. B}\ }\textbf {\bibinfo {volume} {101}},\ \bibinfo
  {pages} {020501} (\bibinfo {year} {2020})}\BibitemShut {NoStop}%
\bibitem [{\citenamefont {Cao}\ \emph {et~al.}(2018)\citenamefont {Cao},
  \citenamefont {Fatemi}, \citenamefont {Demir}, \citenamefont {Fang},
  \citenamefont {Tomarken}, \citenamefont {Luo}, \citenamefont
  {Sanchez-Yamagishi}, \citenamefont {Watanabe}, \citenamefont {Taniguchi},
  \citenamefont {Kaxiras}, \citenamefont {Ashoori},\ and\ \citenamefont
  {Jarillo-Herrero}}]{Cao18a}%
  \BibitemOpen
  \bibfield  {author} {\bibinfo {author} {\bibfnamefont {Y.}~\bibnamefont
  {Cao}}, \bibinfo {author} {\bibfnamefont {V.}~\bibnamefont {Fatemi}},
  \bibinfo {author} {\bibfnamefont {A.}~\bibnamefont {Demir}}, \bibinfo
  {author} {\bibfnamefont {S.}~\bibnamefont {Fang}}, \bibinfo {author}
  {\bibfnamefont {S.~L.}\ \bibnamefont {Tomarken}}, \bibinfo {author}
  {\bibfnamefont {J.~Y.}\ \bibnamefont {Luo}}, \bibinfo {author} {\bibfnamefont
  {J.~D.}\ \bibnamefont {Sanchez-Yamagishi}}, \bibinfo {author} {\bibfnamefont
  {K.}~\bibnamefont {Watanabe}}, \bibinfo {author} {\bibfnamefont
  {T.}~\bibnamefont {Taniguchi}}, \bibinfo {author} {\bibfnamefont
  {E.}~\bibnamefont {Kaxiras}}, \bibinfo {author} {\bibfnamefont {R.~C.}\
  \bibnamefont {Ashoori}}, \ and\ \bibinfo {author} {\bibfnamefont
  {P.}~\bibnamefont {Jarillo-Herrero}},\ }\href {\doibase 10.1038/nature26154}
  {\bibfield  {journal} {\bibinfo  {journal} {Nature}\ }\textbf {\bibinfo
  {volume} {556}},\ \bibinfo {pages} {80} (\bibinfo {year} {2018})}\BibitemShut
  {NoStop}%
\bibitem [{\citenamefont {Gutzwiller}(1963)}]{Gutzwiller63}%
  \BibitemOpen
  \bibfield  {author} {\bibinfo {author} {\bibfnamefont {M.~C.}\ \bibnamefont
  {Gutzwiller}},\ }\href {\doibase 10.1103/PhysRevLett.10.159} {\bibfield
  {journal} {\bibinfo  {journal} {Phys. Rev. Lett.}\ }\textbf {\bibinfo
  {volume} {10}},\ \bibinfo {pages} {159} (\bibinfo {year} {1963})}\BibitemShut
  {NoStop}%
\bibitem [{\citenamefont {Edegger}\ \emph {et~al.}(2007)\citenamefont
  {Edegger}, \citenamefont {Muthukumar},\ and\ \citenamefont
  {Gros}}]{Edegger07}%
  \BibitemOpen
  \bibfield  {author} {\bibinfo {author} {\bibfnamefont {B.}~\bibnamefont
  {Edegger}}, \bibinfo {author} {\bibfnamefont {V.~N.}\ \bibnamefont
  {Muthukumar}}, \ and\ \bibinfo {author} {\bibfnamefont {C.}~\bibnamefont
  {Gros}},\ }\href {\doibase 10.1080/00018730701627707} {\bibfield  {journal}
  {\bibinfo  {journal} {Advances in Physics}\ }\textbf {\bibinfo {volume}
  {56}},\ \bibinfo {pages} {927} (\bibinfo {year} {2007})}\BibitemShut
  {NoStop}%
\bibitem [{\citenamefont {Ko}\ \emph {et~al.}(2007)\citenamefont {Ko},
  \citenamefont {Nave},\ and\ \citenamefont {Lee}}]{Ko07}%
  \BibitemOpen
  \bibfield  {author} {\bibinfo {author} {\bibfnamefont {W.-H.}\ \bibnamefont
  {Ko}}, \bibinfo {author} {\bibfnamefont {C.~P.}\ \bibnamefont {Nave}}, \ and\
  \bibinfo {author} {\bibfnamefont {P.~A.}\ \bibnamefont {Lee}},\ }\href
  {\doibase 10.1103/PhysRevB.76.245113} {\bibfield  {journal} {\bibinfo
  {journal} {Phys. Rev. B}\ }\textbf {\bibinfo {volume} {76}},\ \bibinfo
  {pages} {245113} (\bibinfo {year} {2007})}\BibitemShut {NoStop}%
\bibitem [{\citenamefont {Wang}\ \emph {et~al.}(2006)\citenamefont {Wang},
  \citenamefont {Wang}, \citenamefont {Chen},\ and\ \citenamefont
  {Zhang}}]{Wang06}%
  \BibitemOpen
  \bibfield  {author} {\bibinfo {author} {\bibfnamefont {Q.-H.}\ \bibnamefont
  {Wang}}, \bibinfo {author} {\bibfnamefont {Z.~D.}\ \bibnamefont {Wang}},
  \bibinfo {author} {\bibfnamefont {Y.}~\bibnamefont {Chen}}, \ and\ \bibinfo
  {author} {\bibfnamefont {F.~C.}\ \bibnamefont {Zhang}},\ }\href {\doibase
  10.1103/PhysRevB.73.092507} {\bibfield  {journal} {\bibinfo  {journal} {Phys.
  Rev. B}\ }\textbf {\bibinfo {volume} {73}},\ \bibinfo {pages} {092507}
  (\bibinfo {year} {2006})}\BibitemShut {NoStop}%
\bibitem [{\citenamefont {Christensen}\ \emph {et~al.}(2011)\citenamefont
  {Christensen}, \citenamefont {Hirschfeld},\ and\ \citenamefont
  {Andersen}}]{Christensen11}%
  \BibitemOpen
  \bibfield  {author} {\bibinfo {author} {\bibfnamefont {R.~B.}\ \bibnamefont
  {Christensen}}, \bibinfo {author} {\bibfnamefont {P.~J.}\ \bibnamefont
  {Hirschfeld}}, \ and\ \bibinfo {author} {\bibfnamefont {B.~M.}\ \bibnamefont
  {Andersen}},\ }\href {\doibase 10.1103/PhysRevB.84.184511} {\bibfield
  {journal} {\bibinfo  {journal} {Phys. Rev. B}\ }\textbf {\bibinfo {volume}
  {84}},\ \bibinfo {pages} {184511} (\bibinfo {year} {2011})}\BibitemShut
  {NoStop}%
\bibitem [{\citenamefont {Yang}\ \emph {et~al.}(2009)\citenamefont {Yang},
  \citenamefont {Chen}, \citenamefont {Rice}, \citenamefont {Sigrist},\ and\
  \citenamefont {Zhang}}]{Yang09}%
  \BibitemOpen
  \bibfield  {author} {\bibinfo {author} {\bibfnamefont {K.-Y.}\ \bibnamefont
  {Yang}}, \bibinfo {author} {\bibfnamefont {W.~Q.}\ \bibnamefont {Chen}},
  \bibinfo {author} {\bibfnamefont {T.~M.}\ \bibnamefont {Rice}}, \bibinfo
  {author} {\bibfnamefont {M.}~\bibnamefont {Sigrist}}, \ and\ \bibinfo
  {author} {\bibfnamefont {F.-C.}\ \bibnamefont {Zhang}},\ }\href {\doibase
  10.1088/1367-2630/11/5/055053} {\bibfield  {journal} {\bibinfo  {journal}
  {New J. Phys.}\ }\textbf {\bibinfo {volume} {11}},\ \bibinfo {pages} {055053}
  (\bibinfo {year} {2009})}\BibitemShut {NoStop}%
\bibitem [{\citenamefont {Chakraborty}\ \emph {et~al.}(2017)\citenamefont
  {Chakraborty}, \citenamefont {Sensarma},\ and\ \citenamefont
  {Ghosal}}]{Chakraborty17a}%
  \BibitemOpen
  \bibfield  {author} {\bibinfo {author} {\bibfnamefont {D.}~\bibnamefont
  {Chakraborty}}, \bibinfo {author} {\bibfnamefont {R.}~\bibnamefont
  {Sensarma}}, \ and\ \bibinfo {author} {\bibfnamefont {A.}~\bibnamefont
  {Ghosal}},\ }\href {\doibase 10.1103/PhysRevB.95.014516} {\bibfield
  {journal} {\bibinfo  {journal} {Phys. Rev. B}\ }\textbf {\bibinfo {volume}
  {95}},\ \bibinfo {pages} {014516} (\bibinfo {year} {2017})}\BibitemShut
  {NoStop}%
\bibitem [{\citenamefont {Norman}(2007)}]{Norman07}%
  \BibitemOpen
  \bibfield  {author} {\bibinfo {author} {\bibfnamefont {M.~R.}\ \bibnamefont
  {Norman}},\ }\href {\doibase 10.1103/PhysRevB.75.184514} {\bibfield
  {journal} {\bibinfo  {journal} {Phys. Rev. B}\ }\textbf {\bibinfo {volume}
  {75}},\ \bibinfo {pages} {184514} (\bibinfo {year} {2007})}\BibitemShut
  {NoStop}%
\bibitem [{\citenamefont {Lyons}\ \emph {et~al.}(1988)\citenamefont {Lyons},
  \citenamefont {Fleury}, \citenamefont {Schneemeyer},\ and\ \citenamefont
  {Waszczak}}]{Lyons88}%
  \BibitemOpen
  \bibfield  {author} {\bibinfo {author} {\bibfnamefont {K.~B.}\ \bibnamefont
  {Lyons}}, \bibinfo {author} {\bibfnamefont {P.~A.}\ \bibnamefont {Fleury}},
  \bibinfo {author} {\bibfnamefont {L.~F.}\ \bibnamefont {Schneemeyer}}, \ and\
  \bibinfo {author} {\bibfnamefont {J.~V.}\ \bibnamefont {Waszczak}},\ }\href
  {\doibase 10.1103/PhysRevLett.60.732} {\bibfield  {journal} {\bibinfo
  {journal} {Phys. Rev. Lett.}\ }\textbf {\bibinfo {volume} {60}},\ \bibinfo
  {pages} {732} (\bibinfo {year} {1988})}\BibitemShut {NoStop}%
\bibitem [{\citenamefont {Tranquada}\ \emph {et~al.}(1989)\citenamefont
  {Tranquada}, \citenamefont {Shirane}, \citenamefont {Keimer}, \citenamefont
  {Shamoto},\ and\ \citenamefont {Sato}}]{Tranquada89}%
  \BibitemOpen
  \bibfield  {author} {\bibinfo {author} {\bibfnamefont {J.~M.}\ \bibnamefont
  {Tranquada}}, \bibinfo {author} {\bibfnamefont {G.}~\bibnamefont {Shirane}},
  \bibinfo {author} {\bibfnamefont {B.}~\bibnamefont {Keimer}}, \bibinfo
  {author} {\bibfnamefont {S.}~\bibnamefont {Shamoto}}, \ and\ \bibinfo
  {author} {\bibfnamefont {M.}~\bibnamefont {Sato}},\ }\href {\doibase
  10.1103/PhysRevB.40.4503} {\bibfield  {journal} {\bibinfo  {journal} {Phys.
  Rev. B}\ }\textbf {\bibinfo {volume} {40}},\ \bibinfo {pages} {4503}
  (\bibinfo {year} {1989})}\BibitemShut {NoStop}%
\bibitem [{\citenamefont {Covaci}\ and\ \citenamefont
  {Marsiglio}(2006)}]{Covaci06}%
  \BibitemOpen
  \bibfield  {author} {\bibinfo {author} {\bibfnamefont {L.}~\bibnamefont
  {Covaci}}\ and\ \bibinfo {author} {\bibfnamefont {F.}~\bibnamefont
  {Marsiglio}},\ }\href {\doibase 10.1103/PhysRevB.73.014503} {\bibfield
  {journal} {\bibinfo  {journal} {Phys. Rev. B}\ }\textbf {\bibinfo {volume}
  {73}},\ \bibinfo {pages} {014503} (\bibinfo {year} {2006})}\BibitemShut
  {NoStop}%
\bibitem [{\citenamefont {Black-Schaffer}\ and\ \citenamefont
  {Doniach}(2008)}]{BlackSchaffer07}%
  \BibitemOpen
  \bibfield  {author} {\bibinfo {author} {\bibfnamefont {A.~M.}\ \bibnamefont
  {Black-Schaffer}}\ and\ \bibinfo {author} {\bibfnamefont {S.}~\bibnamefont
  {Doniach}},\ }\href {\doibase 10.1103/PhysRevB.78.024504} {\bibfield
  {journal} {\bibinfo  {journal} {Phys. Rev. B}\ }\textbf {\bibinfo {volume}
  {78}},\ \bibinfo {pages} {024504} (\bibinfo {year} {2008})}\BibitemShut
  {NoStop}%
\bibitem [{\citenamefont {Wang}\ and\ \citenamefont {Lee}(2012)}]{Wang12}%
  \BibitemOpen
  \bibfield  {author} {\bibinfo {author} {\bibfnamefont {F.}~\bibnamefont
  {Wang}}\ and\ \bibinfo {author} {\bibfnamefont {D.-H.}\ \bibnamefont {Lee}},\
  }\href {\doibase 10.1103/PhysRevB.86.094512} {\bibfield  {journal} {\bibinfo
  {journal} {Phys. Rev. B}\ }\textbf {\bibinfo {volume} {86}},\ \bibinfo
  {pages} {094512} (\bibinfo {year} {2012})}\BibitemShut {NoStop}%
\end{thebibliography}%

\pagebreak
\widetext
\clearpage 
\normalsize
~\vspace{0.2cm} 
\setcounter{equation}{0}
\setcounter{figure}{0}
\setcounter{table}{0}
\setcounter{page}{1}
\makeatletter
\renewcommand{\theequation}{S\arabic{equation}}
\renewcommand{\thefigure}{S\arabic{figure}}
\renewcommand{\figurename}{FIG.}

\begin{center}
{\large\bf Supplementary Material for \\
 ``Disorder-robust phase crystal in high-temperature superconductors stabilized by strong correlations"}
\end{center}

\vspace{0.2cm}

In this Supplementary Material (SM) we provide additional details to further support some of the results in the main text. In particular, we first show the circulating currents near the pair breaking edge obtained within GIMT, then we show the temperature dependence of the low-energy eigenvalues of the ground state for a clean superconductor, then we provide a detailed discussion on the relation between the number of zero-energy states and the local charge density in the uniform phase state of a clean superconductor, and finally we illustrate the disorder robustness of phase crystals in GIMT with a different model for the disorder.

\section{Circulating currents near the pair breaking edge}\label{sec:current}

\begin{figure*}[h]
\includegraphics[width=0.8\linewidth]{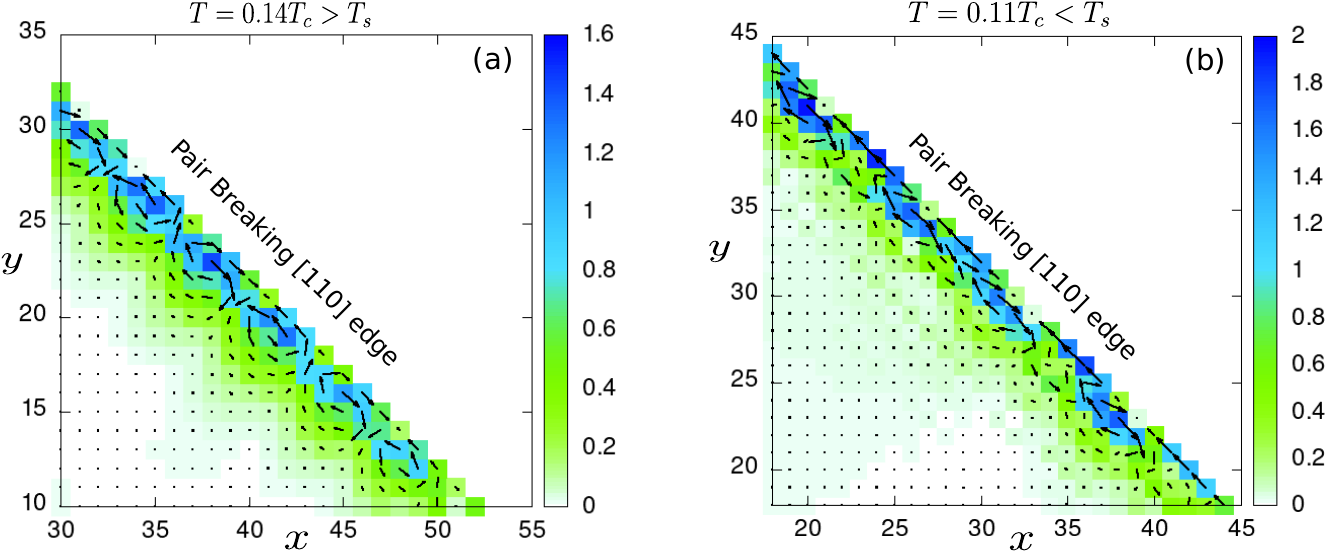}\caption{{\bf{Circulating currents at pair breaking edge.}} Color density maps of the magnitude of the current vector field $\vec{J}_{i}$ and the corresponding vector plots at two different temperatures, (a) in the phase crystal phase and (b) with additional extended $s$-wave pairing, calculated within GIMT. For a better visualization, the currents are scaled by 0.016 and 0.032 in (a) and (b), respectively. Only a part of the whole system is shown here for clarity.}
\label{fig:current} 
\end{figure*}
In the main text, we state that time-reversal symmetry is broken for $T<T^*$ due to the formation of the phase crystal with its modulating phase. In this section, we show that the broken time-reversal symmetry also leads to the appearance of circulating currents. Using the charge continuity equation together with the Heisenberg equation for the particle density (see e.g.~Refs.~\onlinecite{Covaci06,BlackSchaffer07}), the total particle (electron) current on a bond between site $i$ and $i+\delta$ for the Hamiltonian in Eqs.~\eqref{eq:tJ} and \eqref{eq:meanfield2} is given by
\begin{equation}
J_{i,i+\delta}=-it_{i,i+\delta}g^{t}_{i,i+\delta}\sum_{\sigma} \langle c_{i \sigma}^{\dagger}c_{i,i+\delta \sigma} - \textrm{H.c.} \rangle.
\label{eq:cur}
\end{equation}
With $J_{i,i+\delta}$ only present on bonds, we also define for visualization purposes a vector field
\begin{equation}
\vec{J}_{i}=\sum_{\delta}J_{i,i+\delta} \hat{\delta},
\label{eq:cur}
\end{equation}
where $\delta=x,y,y+x,y-x$ and $\hat{\delta}$ is the unit vector along the direction of $\delta$. 
In Fig.~\ref{fig:current} we illustrate the spontaneous currents in a clean sample within GIMT at two different temperatures. Here we use a color density plot for the magnitude of the current vector field $\vec{J}_{i}$ along with a vector plot for the direction of the current vector field. 
As seen in Fig.~\ref{fig:current}(a) for $T_s<T<T^*$, i.e.~in the phase crystal state but without extended $s$-wave pairing, we find alternating circulating currents near the pair breaking edge. This result is very similar to the findings of quasi-classical theory \cite{Hakansson2015,Holmvall20} and BdG \cite{Wennerdal20} for the phase crystal state.
However, for $T<T_s$, where extended $s$-wave pairing also appears near the nodes of the phase crystal, this pictures changes. Within this temperature regime, local $d+is$-wave regions are formed and they add additional currents \cite{Black-Schaffer13} on top of the circulating currents from the phase crystal. The result is the current pattern in Fig.~\ref{fig:current}(b), where the current is clearly finite but does not show a simple circulating pattern.

\section{Temperature dependence of the low-energy eigenvalues}\label{sec:Tdep}

\begin{figure*}[h]
\includegraphics[width=0.8\linewidth]{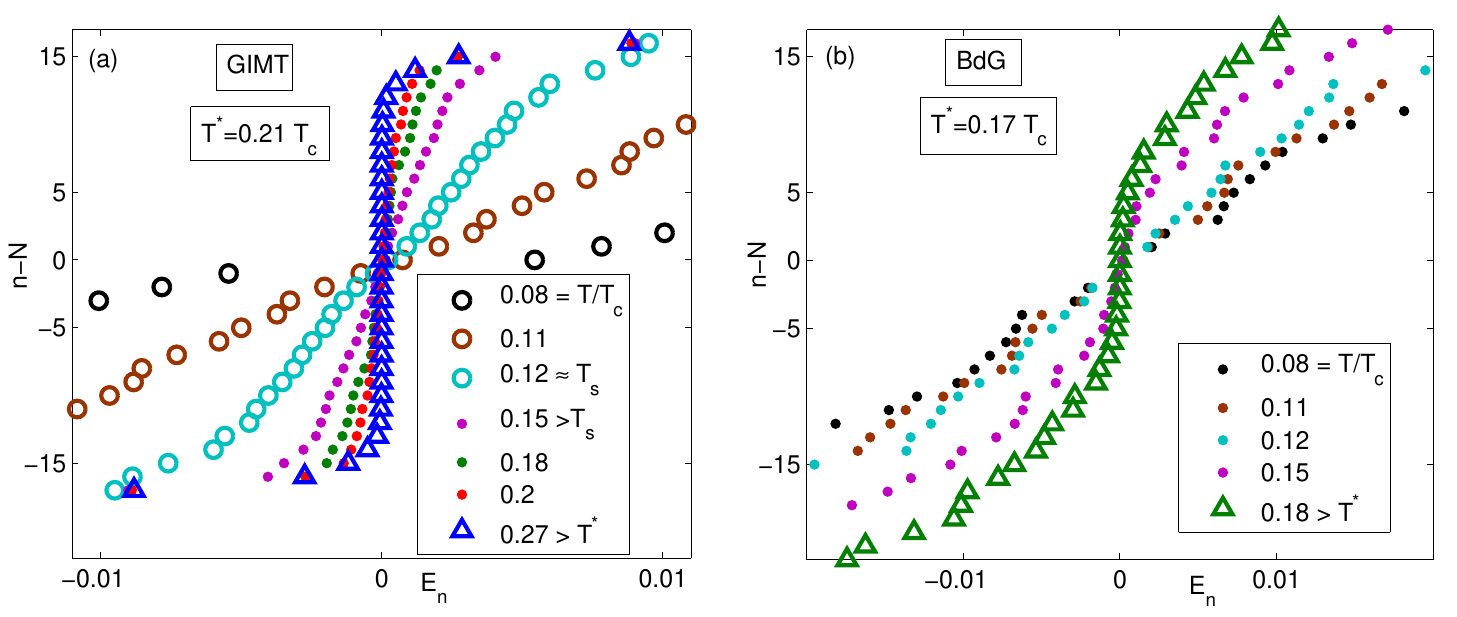}\caption{{\bf{Temperature dependence of the distribution of low-energy eigenvalues for clean sample.}} The temperature evolution of $E_n$ of the ground state (the state with lowest free energy: the phase crystal state for $T<T^*$ and the uniform phase state for $T>T^*$) are shown for GIMT ({\bf{a}}) and BdG ({\bf{b}}). The full gap in GIMT disappears for $T>T_s$ with the slope of the curve increasing very slowly as $T^*$ is approached. In contrast, the slopes of the curves change monotonically with temperature in BdG. Note that $E_n$ for $T/T_c=0.2$ and $T/T_c=0.27$ are not shown in ({\bf{b}}), as the distribution does not change with temperature for $T^*<T<T_c$.}
\label{fig:Tdep} 
\end{figure*}

In Fig.~\ref{fig:thermoplot2} of the main text we show the distribution of the low-energy eigenvalues of a clean sample but only for one specific low temperature. To offer complementary data, we plot in Fig.~\ref{fig:Tdep} the temperature evolution of the eigenvalues $E_n$ of the ground state, i.e.~the phase crystal state for $T<T^*$ and the uniform phase state for $T>T^*$. Within the GIMT, where strong correlation effects are included, we see in Fig.~\ref{fig:Tdep}(a) that there are two temperature regimes for $T<T^*$. First, for $T\le T_s$, where $T_s$ marks the emergence of an extended $s$-wave component at the edge, the full gap in $E_n$ around zero energy reduces with increased temperature. Furthermore, with increasing temperature, more states move toward zero energy, as seen from the increased curve slopes. The rate of slope increase is strong in this temperature regime. Secondly, for $T_s<T\le T^*$ the rate at which the states move towards zero energy is much slower with increasing temperature and eventually the distribution $E_n$ reaches that of the uniform phase state found at $T=T^*$. We note here that these temperature dependencies found below $T^*$ are affecting the eigenvalues of the system, and are as such a temperature effect that goes beyond the standard thermal broadening of physical properties caused by the Fermi-Dirac distribution of electrons. As such, they will cause temperature effects, such as conductance peak broadening, that cannot be related to standard Fermi-Dirac temperature broadening.
 In the uniform phase state the low-energy $E_n$ are instead insensitive to temperature. 
In Fig.~\ref{fig:Tdep}(b) we present the corresponding plot within BdG where strong correlations effects are ignored. Here, the rate at which the $E_n$ distribution approaches the uniform phase state is essentially monotonic in temperature and there is also no emerging $s$-wave component leaving the spectrum ungapped at all temperatures. This contrasting temperature evolution of the eigenvalues is also responsible for a larger $T^*$ in GIMT.

\section{Relation between number of zero-energy states and local charge density}\label{sec:ZESdens}

\begin{figure*}[h]
\includegraphics[width=0.8\linewidth]{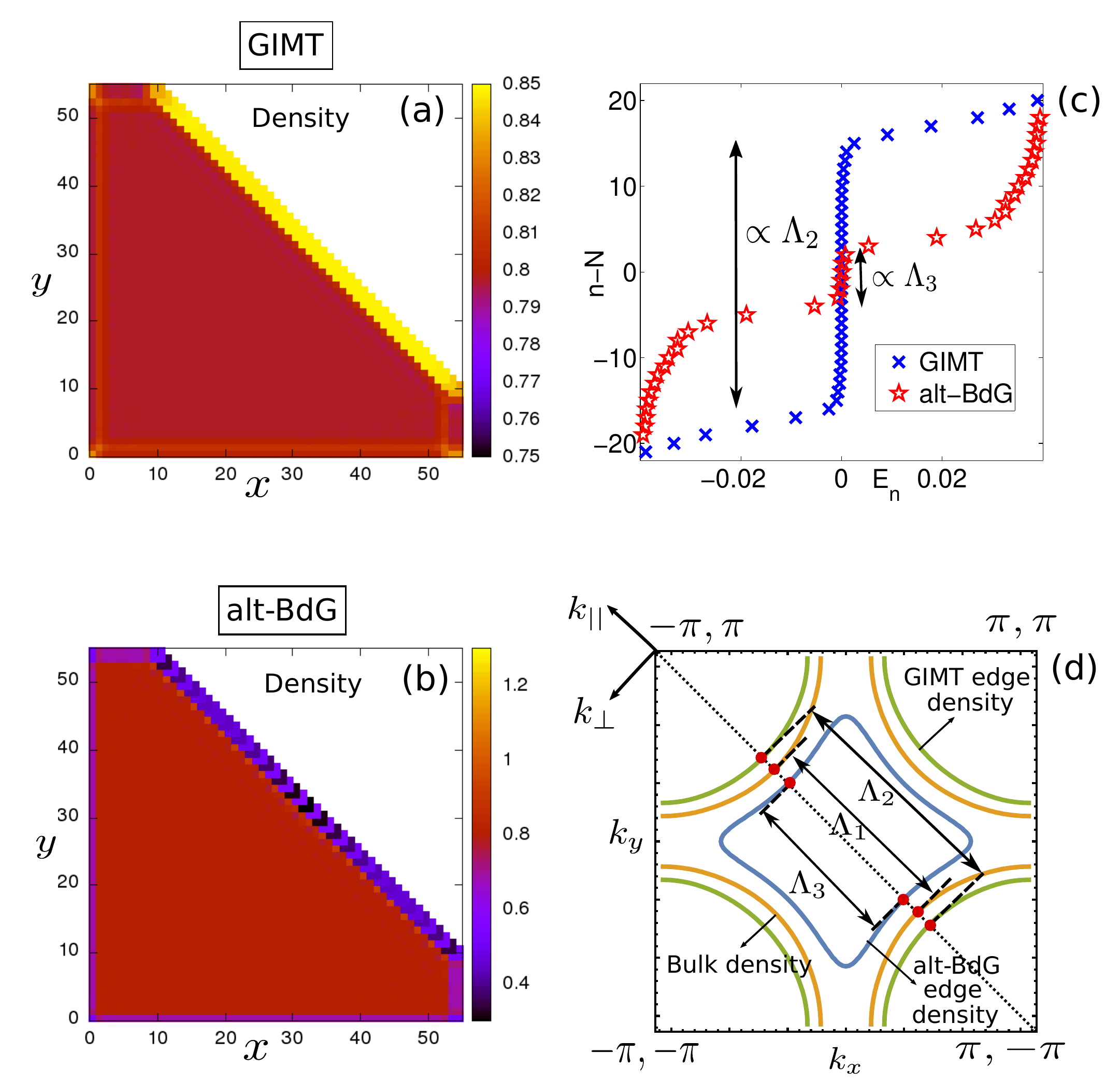}\caption{{\bf{Role of local density for clean sample.}} Spatial color maps of the charge density $\rho_i$ in GIMT ({\bf{a}}) and alt-BdG ({\bf{b}}) for the uniform phase state at a temperature $T=0.08T_c$ in a clean superconductor. Note the exact same bulk charge density (and same pairing amplitude). ({\bf{c}}) Distribution of low-energy eigenvalues $E_n$ of GIMT and alt-BdG. ({\bf{d}}) Fermi surface of bands with homogeneous $\mu$ fixed such that homogeneous charge density is $\rho=0.8$ (bulk), $\rho=0.85$ (edge density in GIMT) and $\rho=0.45$ (edge density in alt-BdG). Dotted line indicates the nodal $k_x=-k_y$ line of the bulk $d$-wave superconductor, red dots the superconducting nodal points, and double black arrows with $\Lambda_i$ labels the distance between superconducting nodal points that sets the lateral extent of the flat band of zero-energy edge states.}
\label{fig:ZESdens} 
\end{figure*}

In this section, we discuss in more detail the reasons behind the increased number of zero-energy states at the edge in the presence of strong correlations, as shown in Fig.~\ref{fig:thermoplot2}(a) of the main text, and why an application of the bulk-boundary correspondence in nodal superconductors using bulk properties is unattainable when strong correlations effects are included.

Strong electron correlations modify the properties of a system in primarily two ways. First, strong correlations renormalize the hopping amplitudes and the super-exchange interaction through the Gutzwiller factors, even in the absence of any inhomogeneity. Secondly, these renormalization factors, $g^t_{ij}$ and $g^{J}_{ij}$, depend on the local charge density and consequently alter the effects of any inhomogeneity.
In order to probe the importance of strong correlations for the zero-energy edge states, we need to focus on the inhomogeneous effects caused by the edge. We can do this by comparing our strongly correlated results with an alternative BdG (alt-BdG) calculation where we choose  parameters such that the bulk solution in the absence of any disorder is exactly the same as in GIMT. Thus, in the alt-BdG calculation we choose $t$, $t^{\prime}$, $J$ and $\mu$ such that the bulk homogeneous solution, both charge density and superconducting pairing amplitude, of Eq.~\eqref{eq:meanfield2} and Eq.~\eqref{eq:BdGH} match exactly. 
In Fig.~\ref{fig:ZESdens}(a,b) we show the resulting spatial color maps of the local charge density $\rho_i$ in the uniform phase state in a clean sample obtained in GIMT and alt-BdG, respectively. By construction, the bulk charge densities are exactly the same in both plots. However, the edge density shows contrasting features: within GIMT $\rho_{\rm edge}>\rho_{\rm bulk}$, while within alt-BdG $\rho_{\rm edge}<\rho_{\rm bulk}$. 
The increase in the charge density at the edge in GIMT is mainly due to the local Hartree shift $\mu_i^{\rm HS}$ given in Eq.~\eqref{eq:hFshifts}. The most significant contribution to $\mu_i^{\rm HS}$ comes from the derivatives of $g^t_{i,i+\delta}$ on the nearest neighbor bonds, i.e.~the first term in Eq.~\eqref{eq:hFshifts}. Since $\partial g^{t}_{i,i+\delta}/\partial \rho_{i}<0$ and the sites on the edge have only two nearest neighbors, $\mu_{\rm edge}^{\rm HS} < \mu_{\rm bulk}^{\rm HS}$. As a consequence, within GIMT edges very generally host a larger charge density than the bulk, not just for the system parameters used to produce Fig.~\ref{fig:ZESdens}.
In contrast, the charge density at system edges in (alt-)BdG, where strong correlation effects are ignored, is either lower (as in Fig.~\ref{fig:ZESdens}(b)) or very similar to the bulk density (as in Fig.~\ref{fig:cleanPC}(d)), depending on the bulk band structure. To summarize, the charge density behavior at the edges in Fig.~\ref{fig:ZESdens}(a,b) is very generic for GIMT and BdG solutions, respectively.

Now that we have fixed the bulk to be the same in GIMT and alt-BdG, we can directly compare the number of zero-energy edge states between the two methods and check how they compare and also how that compares with the prediction of the bulk-boundary correspondence. In Fig.~\ref{fig:ZESdens}(c) we plot the distribution of the low-energy eigenvalues for the uniform phase state in a clean superconductor. We find that the number of zero-energy states is much larger in GIMT than in alt-BdG, quite similar to the results in the main text where we instead compared GIMT to BdG. Most notably, we find a very different number of zero-energy states in GIMT and alt-BdG, despite exactly the same bulk properties. Thus we have explicitly shown that applying the bulk-boundary correspondence using the bulk properties cannot possibly generate the correct number of edge states in both cases

To explain this discrepancy between number of edge states and the bulk properties, we plot in Fig.~\ref{fig:ZESdens}(d) the Fermi surface for three different cases: (1) a band giving the bulk charge density (yellow), (2) a band giving a bulk density equal to the edge density of GIMT (green), and (3) a band giving the bulk density equal to the edge density of alt-BdG (blue). 
Due to translational invariance along $x_{||}$, the edge states of the [110] edge form one-dimensional flat bands at zero energy that terminate at the bulk superconducting nodal points \cite{Potter14,Wang12}, indicated for each case by the red dots in Fig.~\ref{fig:ZESdens}(d). The number of zero-energy edge states is thus proportional to the distance between these bulk nodal points, indicated by the black double-arrows and $\Lambda_i$ labels for the three bands $i=1,2,3$. As a consequence, the number of zero-energy edge states is directly related to the normal-state band structure and its corresponding charge density. However, the problem in a real system is to know which charge density to use when it is changing between the edge and the bulk? With the edge states physically most connected to the edge charge density, it is natural to assume that the number of zero-energy edge states is more related to the edge charge density than that of the bulk.
Based on this assumption connecting the edge charge density with number of zero-energy states, the number of zero-energy states becomes proportional to $\Lambda_3$ in alt-BdG, but $\Lambda_2$ in GIMT, as shown in Fig.~\ref{fig:ZESdens}(c,d). Since $\Lambda_2 > \Lambda_3$ and also $\Lambda_2 > \Lambda_1$, we always get more zero-energy states in the GIMT solution than in any BdG solution, even if the bulk conditions are exactly the same, as we also see in our numerical results. We thus conclude that a simple-minded application of the bulk-boundary correspondence in nodal superconductors to extract the number of zero-energy states from the bulk band structure and its charge density is not correct, especially not in the presence of strong correlations where the charge density at the edge always increases, irrespective of the normal state in the bulk. Additionally, our results show that including strong correlations always leads to more zero-energy edge states due to an enhanced edge charge density, which enhances the stability of the phase crystal state in the presence of strong correlations.

\section{Disorder robustness of phase crystal state for concentration disorder}\label{sec:concdisorder}

\begin{figure*}[h]
\includegraphics[width=1.0\linewidth]{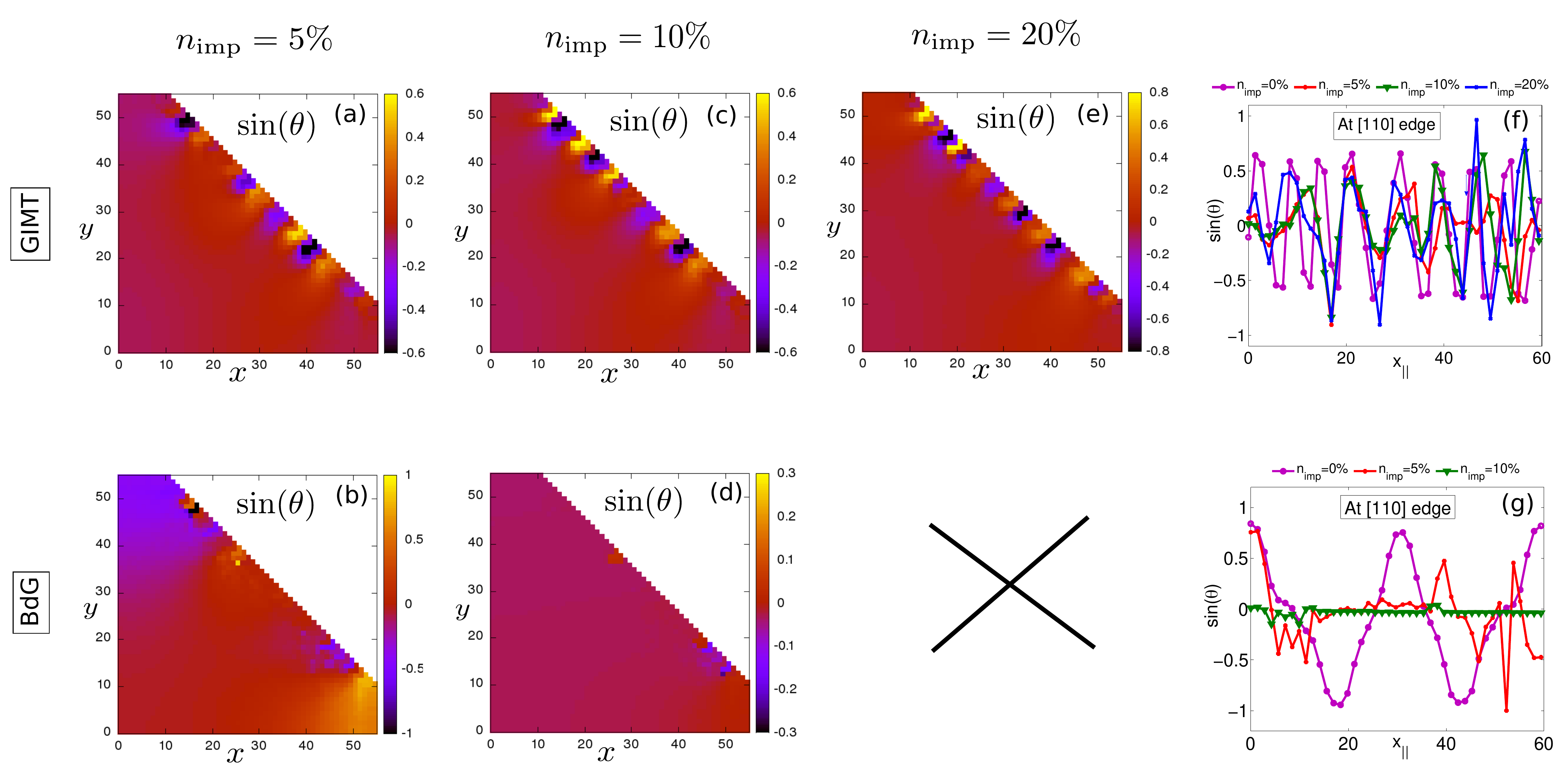}\caption{{\bf{Phase crystals for disordered sample with concentration disorder.}} Color density maps of $\sin(\theta)$ in both GIMT ({\bf{a}},{\bf{c}},{\bf{e}}) and BdG ({\bf{b}},{\bf{d}}) for $T=0.32T^*<T_s$. The phase crystal state does not survive in BdG for $n_{\rm imp}=20\%$ and is hence not plotted. Line plots of the phase modulations at the pair breaking [110] edge as a function of $x_{\parallel}$ are shown for GIMT ({\bf{f}}) and BdG ({\bf{g}}) for different impurity concentrations $n_{\rm imp}$.}
\label{fig:concdisorder} 
\end{figure*}

In the main text, we show the results for `Anderson type' disorder, where a random potential $V_i$ is added to each lattice site. To verify that our results are not sensitive to this particular choice of disorder, we also consider another model of disorder, referred to as concentration disorder, where $V_i=V_0=1$ only on a given $n_{{\rm imp}}$ fraction of randomly chosen lattice sites, while $V_i=0$ for all other sites. Thus concentration disorder contains fewer but stronger scatterers, while Anderson disorder instead generates a continuum of relatively weaker scatterers.
The results when using concentration disorder are shown in Fig.~\ref{fig:concdisorder}, where we plot the color density maps of $\sin(\theta)$ in both GIMT and BdG for different concentrations $n_{\rm imp}$. The remarkable robustness of the phase crystal state in GIMT persists even up to the very high disorder concentration $n_{\rm imp}=20\%$, which only induces minor irregularities in the phase crystal modulations. In contrast, the phase crystal state in BdG is extremely sensitive to concentration disorder, with the modulations getting disrupted even for a low concentration $n_{\rm imp}=5\%$. We thus conclude that the disorder results obtained in the main text are not sensitive to the type of the disorder and the phase crystal state is robust to all types of (non-magnetic) disorder when strong correlations are appropriately included.

\end{document}